\documentclass[aps,prx,reprint,notitlepage,longbibliography,floatfix]{revtex4-2}
\usepackage[T1]{fontenc}
\usepackage{amsmath,amssymb,bm}
\usepackage{graphicx}
\usepackage[colorlinks=true,linkcolor=blue,citecolor=blue,urlcolor=blue]{hyperref}
\newcommand{\Tr}{\operatorname{Tr}}
\newcommand{\ad}{\operatorname{ad}}
\usepackage{graphicx,amsmath,subfigure}%
\usepackage{color,xcolor}

\usepackage{epstopdf}
\usepackage{float}
\usepackage{ulem}
\usepackage{upgreek}
\usepackage{silence}
\usepackage{inputenc}\inputencoding{utf8}\DeclareUnicodeCharacter{200B}{}

\usepackage{booktabs}
\usepackage{amsthm}
\newtheorem{theorem}{Theorem}[section]

\newtheorem{proposition}[theorem]{Proposition}

\theoremstyle{definition}

\theoremstyle{remark}

\newcommand{\E}{\mathbb E}

\begin{document}

\title{Logarithmic Bias of Zero-Noise Extrapolation under \texorpdfstring{$1/f$}{1/f} Noise}
% Fill in the agreed author list, affiliations, and corresponding author
% before public submission. No authorship metadata has been inferred.
\author{Chenhui Wang}
\email{ccchui123@outlook.com}
\affiliation{Laboratory for Advanced Computing and Intelligence Engineering, Zhengzhou 450001, Henan, China}
\author{Weilong Wang} 
\affiliation{Laboratory for Advanced Computing and Intelligence Engineering, Zhengzhou 450001, Henan, China}
\author{Haoran He} 
\affiliation{Laboratory for Advanced Computing and Intelligence Engineering, Zhengzhou 450001, Henan, China}
\author{Ping Lv} 
\affiliation{Laboratory for Advanced Computing and Intelligence Engineering, Zhengzhou 450001, Henan, China}
\author{Zheng Shan}
\email{shanzhengzz@163.com}
\affiliation{Laboratory for Advanced Computing and Intelligence Engineering, Zhengzhou 450001, Henan, China}

\date{October 13, 2026}

\begin{abstract}
Zero-noise extrapolation estimates ideal quantum observables by amplifying
noise and extrapolating the measured response. Its accuracy depends on
whether the extrapolation model captures the physical effect of amplification,
including correlations between repeated operations. Here we derive the
response of complete multiqubit gate folding in the simultaneous presence
of time-local Markovian noise and stationary finite-band $1/f$ noise.
An ordered circuit kernel couples the observable to noise memory; exact
block counting shows how folding samples correlations at increasing time
separations. The resulting response contains an $r^2\ln r$ coordinate,
with a coefficient fixed by the noise spectrum and the integrated
forward--inverse control response. Ordinary polynomial Richardson
extrapolation leaves a calculable contribution from this coordinate;
the three-node rule $(1,3,5)$ retains $2.73\theta$.
We construct response-matched extrapolation, including a log-domain
implementation, that cancels this contribution at the retained order.
Explicit bounds separate the remaining control-shape, finite-band,
higher-order, and native-offset errors. The theory applies to arbitrary
finite multiqubit circuits within the stated control and noise model,
and distinguishes removal of memory-induced bias from an unconditional
guarantee of the ideal intercept. Single- and ten-qubit numerical
examples illustrate the predicted correction and its response dependence.
\end{abstract}
\maketitle
\raggedbottom

\section{Introduction}
\label{sec:introduction}
Zero-noise extrapolation (ZNE) infers ideal expectation values from
a family of logically equivalent circuits executed at amplified noise
levels~\cite{Temme2017ErrorMitigation,LiBenjamin2017ActiveErrorMinimization}.
Pulse stretching changes the duration of a control sequence while
preserving its ideal action, whereas unitary folding inserts
inverse--forward pairs~\cite{Kandala2019ComputationalReach,GiurgicaTiron2020DigitalZNE}.
The attraction is operational simplicity: measured observables, rather
than an encoded quantum state, are corrected. The central tension is
that logical equivalence does not determine the noise dependence of
those observables.

Two obstacles arise when the noise has memory. First, a polynomial or
single-exponential fitting model does not by itself account for the
correlations sampled by the amplification protocol. Second, without a
response calculation, the resulting model bias cannot be predicted
from the noise and control, so the missing extrapolation coordinate
is not identified. These are questions about the population response,
not only the precision with which it is sampled. Time-local noise
models motivate useful analytic expansions, but memoryless noise is
neither a defining requirement of ZNE nor a sufficient condition for
every proposed fit to be accurate.

Low-frequency noise with an approximately $1/f$ spectrum is observed
in several solid-state qubit settings~\cite{Paladino2014OneOverF,Krantz2019SuperconductingGuide}.
Its consequences depend on the coupling and control, and it need not
dominate every error channel. Previous work has analyzed the effect of
time-correlated noise on ZNE and identified differences between local
and global amplification~\cite{Schultz2022TimeCorrelatedZNE}.
ZNE has also been implemented in silicon spin qubits using global
folding~\cite{Sohn2025SiliconZNE}. Separately, reactivity-based methods
describe extrapolation curvature arising from circuit dynamics under
Markovian channel models~\cite{Suchsland2026InformationDynamics}.
The question addressed here is the explicit extrapolation law generated
by temporal memory during finite-duration, complete-circuit folding.

Our first contribution is a response theory that links this memory to
the extrapolation coordinate. At second order in the colored-noise
coupling, an ordered circuit kernel is contracted with the covariance.
An exact second-difference identity isolates correlations between
translated blocks. For finite-band $1/f$ noise this yields a discrete
logarithmic hierarchy and a controlled four-coordinate reduction,
including $r^2\ln r$. Its coefficient is determined before fitting
by the spectrum, control, input state, and observable.
The familiar logarithmic time dependence of $1/f$ dephasing is not
itself new~\cite{Paladino2014OneOverF}; the result here is its consequence
for the folding response and the bias of extrapolation.

Our second contribution is a response-matched correction. We derive the
logarithmic bias component of ordinary Richardson rules and construct
weights that cancel it together with the regular linear and quadratic
coordinates. The resulting log-domain method, PE\_log, uses the same
response coordinates at the retained noise order. A linear-domain
variant admits an explicit error certificate and a conditional
bias-improvement theorem with a nonempty domain. The coefficients,
remainders, and ideal-intercept conditions are kept separate:
correcting the identified coordinate does not remove every source
of extrapolation error.

Section~\ref{sec:response} derives the memory-dependent response.
Section~\ref{sec:bias} identifies its consequence for conventional
extrapolation, and Sec.~\ref{sec:protocol} constructs the correction.
Section~\ref{sec:numerical} presents numerical illustrations, and
Sec.~\ref{sec:discussion} discusses scope and implementation.
The Supplemental Material (SM) contains the proofs, error bounds,
and the comparison with pulse stretching.

% Unmodified artwork: folding_1f_sim/figures/fig1_theory_v2.pdf.
% Numerical source: results/T5_fig1_theory, including its full kernel.
\begin{figure}[t]
\centering
\includegraphics[width=\columnwidth]{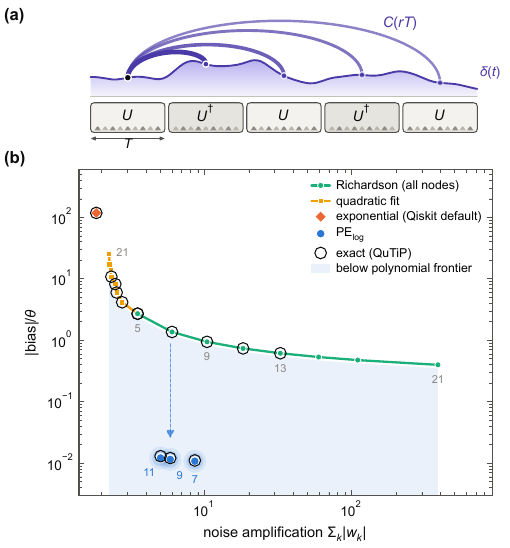}
\caption{\textbf{Folding reads noise memory and exposes bias omitted
by conventional fitting models.}
(a) Correlations between forward and inverse blocks. The arcs
illustrate block-separated memory; Eq.~\eqref{eq:curvature}
integrates delays around $rT$, rather than sampling a single delay.
(b) Normalized predicted bias versus intercept-weight sensitivity
$\sum_k|w_k|$, not the physical folding factor. The shaded
``polynomial frontier'' covers only the displayed quadratic-fit
and Richardson families, not all polynomial estimators. Filled
markers use the second-order kick response; open circles use
untruncated QuTiP propagation at $\chi(13)=10^{-4}$, with finite
ensemble error. Richardson uses $\{1,3,\ldots,r_{\max}\}$;
annotations give $r_{\max}$. Frozen-memory bias vanishes at second
order for degree-two-or-higher polynomial rules under the matching
and survival conditions. Panels (a) and (b) use different infrared
rates. SM, Sec.~S8, specifies both smooth-band models, signal
normalization and estimators. No hardware data are shown.}
\label{fig:theory}
\end{figure}

\section{Noise Memory in Folded Circuits}
\label{sec:response}
\subsection{The response kernel}
Consider a finite multiqubit register, ideal target unitary $U$,
initial state $\rho_0$, and measured Hermitian observable $O$.
Write $O_U=U^\dagger OU$ and $E^*=\Tr(O_U\rho_0)$.
The implemented circuit is subject to a bounded time-local Markovian
generator and a classical perturbation
$\lambda\sum_\alpha\xi_\alpha(t)B_\alpha(t)$.
The fields are centered and stationary, with
$\langle\xi_\alpha(t)\xi_\beta(s)\rangle
=\mathsf A_{\alpha\beta}C(t-s)$, where $\mathsf A$ is a real symmetric
positive-semidefinite spatial covariance matrix. The control and
coupling operators may be multiqubit and need not commute.

For a folded circuit of duration $D_r$, the interaction-picture
couplings $b_{r,\alpha}=U_r^\dagger B_{r,\alpha}U_r$ determine the
ordered kernel
\[
 K_{\alpha\beta,r}(t,s)=
 \Tr(O_U[b_{r,\alpha}(t),[b_{r,\beta}(s),\rho_0]]) .
\]
On a fixed nonzero sign branch with $E^*\ne0$, the central response
relation is
\begin{equation}
\begin{aligned}
 \ln\left|\frac{E(r)}{E^*}\right|
 ={}&\frac{W_r}{E^*}
 -\frac{\lambda^2}{E^*}\sum_{\alpha\beta}\mathsf A_{\alpha\beta}
 \int_0^{D_r}\!dt\int_0^t\!ds\\[-1mm]
 &\hspace{17mm}\times K_{\alpha\beta,r}(t,s)C(t-s)
 +R_{\log}(r).
\end{aligned}
\label{eq:response}
\end{equation}
Here $E(r)$ is the physical ensemble mean and $W_r$ is the first-order
white-noise response. Equation~\eqref{eq:response} follows from the
ordered Dyson expansion followed by a local expansion of the logarithm:
it retains first white-noise order and second colored-noise order.
It is not an exact all-orders cumulant closure. The remainder contains
higher noise orders, mixed terms, and log-transformation corrections
(SM, Secs.~S1 and S7).

The same kernel relation holds in the linear domain without requiring
$E^*\ne0$. Gaussianity is not needed to obtain its second-order
coefficient, but is used in the explicit higher-order bound below.
The kernel is observable-dependent and is not generally a positive
scalar filter power. Equation~\eqref{eq:response} separates two physical
inputs: the circuit weights the time pairs, and the covariance specifies
the memory carried by those pairs.

Complete global folding executes the chronological sequence
$F,I,F,\ldots,F$, with $F$ implementing $U$, $I$ implementing $U^\dagger$,
and $r=2n+1$ equal-duration blocks of duration $T$.
This includes the entire single-qubit and entangling-gate sequence.
Every forward copy repeats the same physical implementation, as does
every inverse copy; the noise realization nevertheless continues
across block boundaries. The common-frame coupling profiles are
\[
\begin{aligned}
 f_\alpha(u)&=U_F^\dagger(u)B_{F,\alpha}(u)U_F(u),\\
 i_\alpha(u)&=U^\dagger U_I^\dagger(u)B_{I,\alpha}(u)U_I(u)U.
\end{aligned}
\]
Logical inversion fixes the endpoints, but does not require these
two profiles to match or to be time reverses. The structural result
below does not impose such a mirror condition.

The first white-noise contribution is affine:
$W_r=a_wr+\Omega$, because there are $(r+1)/2$ forward and
$(r-1)/2$ inverse copies. Colored noise instead couples distinct
blocks through $C(dT+u-v)$, with $d$ their separation in block units.
Define $\Delta_2^2E(r)=E(r+2)-2E(r)+E(r-2)$ for odd $r\geq3$.
Exact counting of all block pairs gives
\begin{equation}
\begin{aligned}
 \Delta_2^2E(r)={}&
 -\lambda^2\int_0^{2T}\!du\int_0^{2T}\!dv\,K_\Delta(u,v)\\
 &\times C(rT+u-v)+\Delta_2^2R_{\rm high}(r).
\end{aligned}
 \label{eq:curvature}
\end{equation}
The fixed kernel $K_\Delta$ is built from two composite common-frame
profiles, $(i,f)$ and $(f,i)$ (SM, Sec.~S3).
Thus curvature samples a weighted range of delays around $rT$,
not a point value $C(rT)$ unless an additional approximation is justified.
The affine white contribution has canceled.

This identity distinguishes memory classes without specifying a
particular circuit. Ideal white memory has no cross-block contribution
at these positive delays. A frozen covariance gives constant curvature
and a quadratic response, possibly with a native constant.
Exponential memory gives a decaying translated-memory contribution.
A logarithmic covariance gives logarithmic curvature. The latter is
the origin of the additional extrapolation coordinate.
Figure~\ref{fig:theory}(a) illustrates the cross-block correlations;
the pulse-pair kernel and numerical memory responses are shown in
SM, Fig.~S1.

\subsection{The logarithmic coordinate}
In the two-sided angular-frequency convention, take
\[
 S_{\alpha\beta}(\omega)=\frac{\mathsf A_{\alpha\beta}}{|\omega|},
 \qquad \omega_{\rm IR}<|\omega|<\omega_{\rm UV},
\]
and zero outside this finite band.
Then
\[
 C(\tau)=
 \frac{\operatorname{Ci}(\omega_{\rm UV}|\tau|)
       -\operatorname{Ci}(\omega_{\rm IR}|\tau|)}{\pi}.
\]
In the intermediate-time window it has logarithmic part
$[\ell_T-\ln(|\tau|/T)]/\pi$,
where $\ell_T=-\gamma_{\rm E}-\ln(\omega_{\rm IR}T)$.
Finite-band corrections are bounded after integration, including the
time diagonal; a pointwise ultraviolet estimate is not integrated
through its singular endpoint.

To see the folding dependence, insert this logarithmic covariance
into Eq.~\eqref{eq:curvature}. For $r\geq3$,
$|(u-v)/(rT)|\leq2/3$, so the expansion of
$\ln[1+(u-v)/(rT)]$ is uniformly convergent.
The second difference of the colored response has the form
$d_0+d_L\ln r+\sum_{p\geq1}d_p/r^p$.
Its discrete integral contains a leading $r^2\ln r$ term and
additional control-shape coordinates. Since
$\Delta_2^2(r^2\ln r)=8\ln r+12+O(r^{-2})$,
the logarithmic coefficient is $d_L/8$.
SM, Sec.~S4, derives the full discrete hierarchy rather than
assuming a four-function ansatz.

Let $Q_{F,\alpha}=\int_0^Tf_\alpha(u)\,du$,
$Q_{I,\alpha}=\int_0^Ti_\alpha(u)\,du$, and
$Q_{+,\alpha}=(Q_{F,\alpha}+Q_{I,\alpha})/2$.
A fixed anchoring of the discrete response gives
\begin{equation}
\begin{aligned}
 E(r)={}&E^*+\delta+a_1r+a_2r^2+\theta r^2\ln r+R(r),\\
 \theta={}&\frac{\lambda^2}{2\pi}
 \sum_{\alpha\beta}\mathsf A_{\alpha\beta}
 \Tr\!\left(O_U[Q_{+,\alpha},[Q_{+,\beta},\rho_0]]\right).
\end{aligned}
\label{eq:fold-response}
\end{equation}
The coefficient is fixed by physical response, not chosen to improve
a fit. Equivalently $\theta=\lambda^2\Gamma_{0,\rm gf}/\pi$,
with $M_0=8\Gamma_{0,\rm gf}$ the integrated curvature kernel.
For independent local noise contributions this is the
spectrum--response contraction
$\theta=\sum_k A_k s_{2,k}/\pi$, with $A_k$ including the coupling
strength and $s_{2,k}$ the corresponding static circuit response.
It may vanish for a noise-insensitive observable without invalidating
the representation.

The remainder in Eq.~\eqref{eq:fold-response} has three explicitly
bounded parts: control shape, finite-band edges, and higher noise orders.
For example, let $\beta_\alpha$ bound the norm of coupling operator
$B_\alpha(t)$ throughout both block implementations. With
$\mathcal B_A=\sum_{\alpha\beta}|\mathsf A_{\alpha\beta}|
\beta_\alpha\beta_\beta$
and $h_{\rm bd}=\lambda^2\|O\|_\infty\mathcal B_AT^2/\pi$,
the generic shape bound grows as $O(h_{\rm bd}r\ln r)$.
This is still colored-noise second order. Mirror matching improves
the shape estimate under a consistently changed anchoring, but is
not required for Eq.~\eqref{eq:fold-response}.
The Gaussian mixed-order bound is
$\|O\|_\infty[e^{K_w+Z}-1-K_w-Z]$, with
$K_w=\int_0^{rT}\|\mathcal L_w^I(t)\|_{1\to1}\,dt$ and
$Z=2\lambda^2(rT)^2
[\sum_\alpha\beta_\alpha\sqrt{C_{\alpha\alpha}(0)}]^2$.
These bounds specify when reduction is controlled; they do not
assume it is always accurate.

The four coordinates are consequently a leading response model,
not an exact finite-dimensional closure. In the stated anchoring
$a_2=-\theta(\ell_T+3/2)$ for the logarithmic colored component;
the fit need not impose this relation or know the infrared cutoff.
Even in the absence of white noise, the folding linear coefficient
need not vanish. More generally, the omitted shape coordinates must
not be reclassified as higher noise orders simply because they are
subleading in $r$.

Pulse stretching reads the same covariance differently. Under the
complete-control rescaling $H_c(t)=H_0(t/c)/c$ and the corresponding
coupling covariance, its response is proportional to $c^2C(c\tau)$.
This yields $c^2\ln c$ with a protocol-specific coefficient and an
exact zero-duration anchor. SM, Secs.~S2, S4, and S6, gives the
full derivation and the conditions under which the two logarithmic
coefficients agree. Their shared coordinate does not make their
constant terms interchangeable. For fixed finite spectral cutoffs,
the stretching response is regular as $c\to0$; its finite-window
logarithmic decomposition must not be extended to claim a physical
singularity at zero.

\section{Bias of Conventional Extrapolation}
\label{sec:bias}
Polynomial Richardson extrapolation cancels polynomial response
coordinates. A single-exponential model instead assumes an affine
log response, after any independently specified offset treatment.
The latter is exact in certain matched Clifford--Pauli channel
models, not for general Markovian dynamics. At the retained order,
a frozen field already generates a quadratic log response, while
$1/f$ memory generates the additional logarithmic coordinate.
Neither a finite polynomial as a function of $r$ nor an affine
function of $r$ in the log domain represents $r^2\ln r$.

For an ordinary degree-$p$ Richardson rule, $p\geq2$, the weights
$g_k$ obey $\sum_kg_k=1$ and $\sum_kg_kr_k^m=0$ for $1\leq m\leq p$.
Substitution of Eq.~\eqref{eq:fold-response} yields
\begin{equation}
\begin{aligned}
 \mathbb E[\widehat E_R]-E^*
 &=\delta+\theta M_{2\log}+\sum_kg_kR(r_k),\\
 M_{2\log}&=\sum_kg_kr_k^2\ln r_k .
\end{aligned}
\label{eq:richardson-bias}
\end{equation}
The isolated logarithmic contribution is therefore calculable from
the node geometry and the forward response coefficient.
For the minimal rule on $p+1$ distinct positive nodes,
\[
 M_{2\log}=
 \int_0^\infty
 \frac{t\prod_{j=0}^{p}r_j}{\prod_{j=0}^{p}(t+r_j)}\,dt>0 .
\]
The integral converges for $p\geq2$. This establishes a nonzero
logarithmic moment for every such ordinary interpolation rule,
not for arbitrary overdetermined or response-matched weights.
On $(1,3,5)$ the weights are $(15/8,-5/4,3/8)$ and the moment
is $(75/8)\ln5-(45/4)\ln3\simeq2.73$ (SM, Sec.~S5).

At fixed nodes, increasing the number of measurements cannot change
this population-level component. This is the precise sense in which
the logarithmic bias is locked to the model. It is not an unconditional
lower bound on the total error: the native offset and remainder in
Eq.~\eqref{eq:richardson-bias} can reinforce or oppose it. Nor does
the result imply a node-independent floor for every growing grid.
The finite-family comparison in Fig.~\ref{fig:theory}(b) illustrates
this distinction: higher-degree interpolation can reduce bias while
increasing weight amplification. Those curves include the full
second-order kernel response and are not just the isolated
$\theta M_{2\log}$ term.

The corresponding mismatch of a single exponential is also
structural: its logarithm has no quadratic or quadratic-logarithmic
coordinate. The theory does not rank exponential extrapolation as
the worst estimator for every circuit or fitting window. It identifies
the memory contribution that its function class omits.

\section{Response-Matched Extrapolation}
\label{sec:protocol}
The kernel calculation supplies the missing coordinate before the
data are fitted. Define the response design
$X_{k\cdot}=(1,r_k,r_k^2,r_k^2\ln r_k)$ and take the intercept weights
$\boldsymbol w^{\mathsf T}=\boldsymbol e_1^{\mathsf T}X^+$.
They satisfy
$\sum_kw_k=1$ and
$\sum_kw_kr_k=\sum_kw_kr_k^2=\sum_kw_kr_k^2\ln r_k=0$.
At least four distinct positive odd fold nodes give full column rank
(SM, Sec.~S7); no particular four nodes are mandatory.

For positive inputs $y_k$ estimating $E(r_k)$, the log-domain
implementation of response-matched PE-ZNE is
\begin{equation}
 \widehat E_{\mathrm{PE,log}}
 =\exp\!\left(\sum_kw_k\ln y_k\right).
 \label{eq:pe-log}
\end{equation}
It is the intercept of an unweighted fit of $\ln y_k$ to the
four response coordinates. The weights depend only on the nodes.
Neither $E^*$ nor $\theta$ is needed to apply them.
Independent knowledge of $\theta$ instead permits a forward test
of the memory mechanism and the predicted conventional bias.

The retained-order justification follows from
$\ln|E|=\ln|E^*|+(E-E^*)/E^*+O((E-E^*)^2)$.
First white-noise and second colored-noise corrections therefore
have the same coordinate set in both domains.
A true exponential attenuation is linearized by the logarithm.
This is a reason to use PE\_log, not a proof that a log transform
removes mixed-noise remainders or restores an ideal anchor.
Zero or sign-changing inputs fall outside the positive-branch
definition, and nearly zero inputs make it unstable.

The linear-domain variant
$\widehat E_{\mathrm{PE,lin}}=\sum_kw_k\widehat E(r_k)$ has a separate
exact expectation-value certificate:
\begin{equation}
\begin{aligned}
 \mathbb E[\widehat E_{\mathrm{PE,lin}}]-E^*
 &=\delta+\sum_kw_kR(r_k),\\
 \left|\mathbb E[\widehat E_{\mathrm{PE,lin}}]-E^*\right|
 &\leq H_{\rm nat}+\|\boldsymbol w\|_1D_{\max}.
\end{aligned}
\label{eq:certificate}
\end{equation}
Here $|\delta|\leq H_{\rm nat}$ and
$D_{\max}=\max_kD(r_k)$ bounds the physical remainders.
The logarithmic coordinate has disappeared without knowing its
coefficient. Equation~\eqref{eq:certificate} does not apply unchanged
to the exponential back-transformation in Eq.~\eqref{eq:pe-log}.

The same certificate gives a concrete bias-improvement condition.
On four nodes $L(1,3,5,7)$, let $\boldsymbol q$ be the quadratic
least-squares intercept weights on those same nodes and let
$\beta=\sum_kq_ks_k^2\ln s_k>0$, with $s_k=1,3,5,7$.
Then
\[
 |\theta|\beta L^2>
 2H_{\rm nat}+(\|\boldsymbol w\|_1+\|\boldsymbol q\|_1)D_{\max}
\]
guarantees a smaller absolute bias for PE\_lin.
SM, Sec.~S7, proves the condition and constructs a nonempty physical
parameter family satisfying it, including a finite-duration entangling
circuit with nonzero response.
This comparison concerns deterministic bias against a specified
quadratic rule, not PE\_log performance or finite-sample optimality.

\paragraph*{The ideal intercept.}
Full folding preserves the ideal unitary at every physical odd node,
but $r=0$ is not a physical zero-duration member of that family.
The inferred constant may contain native noise contributions.
Matching the forward and inverse white-noise responses removes the
first-order white constant $\Omega$. Under colored-response mirror
matching and $[\rho_0,O_U]=0$, the infrared-enhanced constant
proportional to $\ell_Ts_0$ also vanishes.
These are distinct conditions. They do not remove the remaining
shape constant, finite-band ultraviolet contribution, or allowed
higher-order constants. SM, Sec.~S6, gives mismatch bounds and a
model-assisted identifiability criterion for more general observables.
A logical inverse, including a self-inverse CZ, does not alone certify
physical matching.

% Queue the wide figures ahead of their discussion to keep them nearby.
% Unmodified artwork; sources: T3 D/F, T8 A, T10.
\begin{figure*}[t]
\centering
\includegraphics[width=7in]{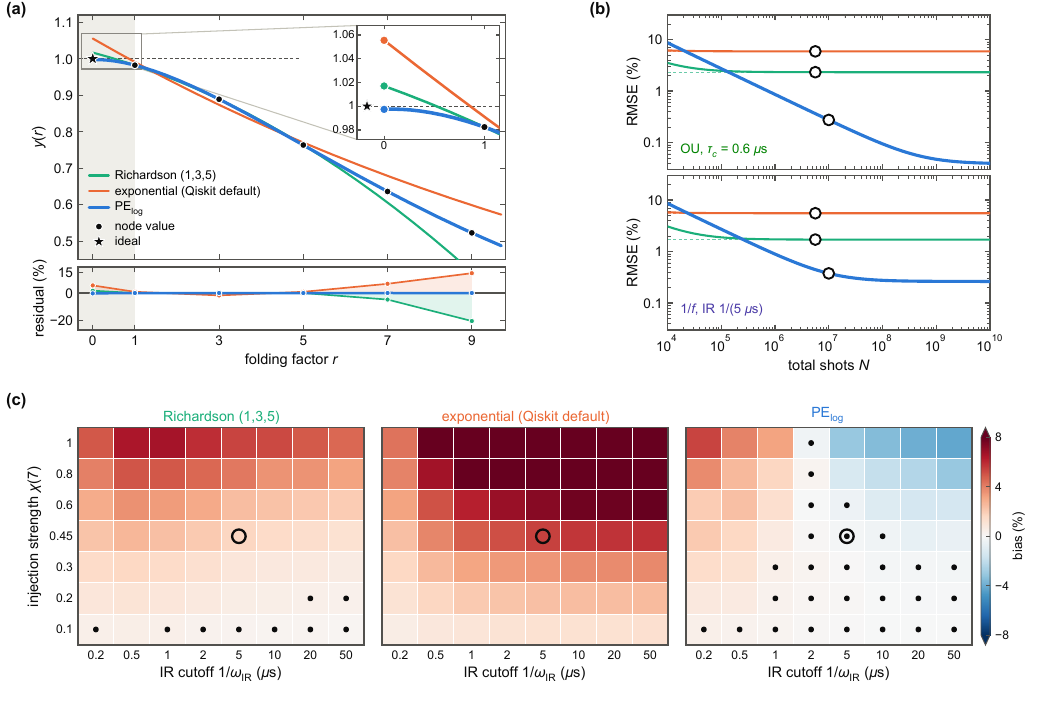}
\caption{\textbf{The memory-matched basis reduces extrapolation bias
in the displayed single-qubit models.} Numerical predictions for
C8 at $\chi(7)=0.45$, except where varied in (c).
(a) A $1/f$ response with inverse infrared rate $5\,\mu\mathrm{s}$,
the three extrapolations, an intercept inset and a residual strip.
Black nodes are computed means, not measured points. The residual is
$100[\widehat y(r)/y(r)-1]$ at a fold node and
$100[\widehat y(0)-1]$ at the intercept. Gray shading marks $r<1$;
colored residual fills are not confidence bands.
Residuals outside a fitted node subset test continuation, whereas
zero interpolation residuals at fitted nodes are not validation.
Bars in (a) project one sampling standard deviation; numerical
ensemble errors are separate.
(b) Projected RMSE, including on/off sampling, for OU memory
($\tau_c=0.6\,\mu\mathrm{s}$) and the $1/f$ setting in (a).
Open markers use equal per-node budgets, not equal total budgets
across the three- and five-node estimators. At $10^7$ total samples,
PE\_log has projected RMSE $0.38\%$ for $1/f$ and $0.27\%$ for OU;
these are not the sampling standard deviations. The nonzero
PE\_log bias also sets a residual floor.
(c) Computed bias versus injection strength and inverse infrared
rate; circles mark the reference setting and dots mark
$|\mathrm{bias}|\leq0.5\%$. The legend's ``Qiskit default'' labels
the exponential fitting form at $(1,3,5)$, not a simulation of the
complete Runtime mitigation workflow. Models and uncertainty
conventions are specified in SM, Sec.~S8.}
\label{fig:correction}
\end{figure*}

% Unmodified artwork; sources: T2 and T11 collapse summary.
\begin{figure*}[t]
\centering
\includegraphics[width=7in]{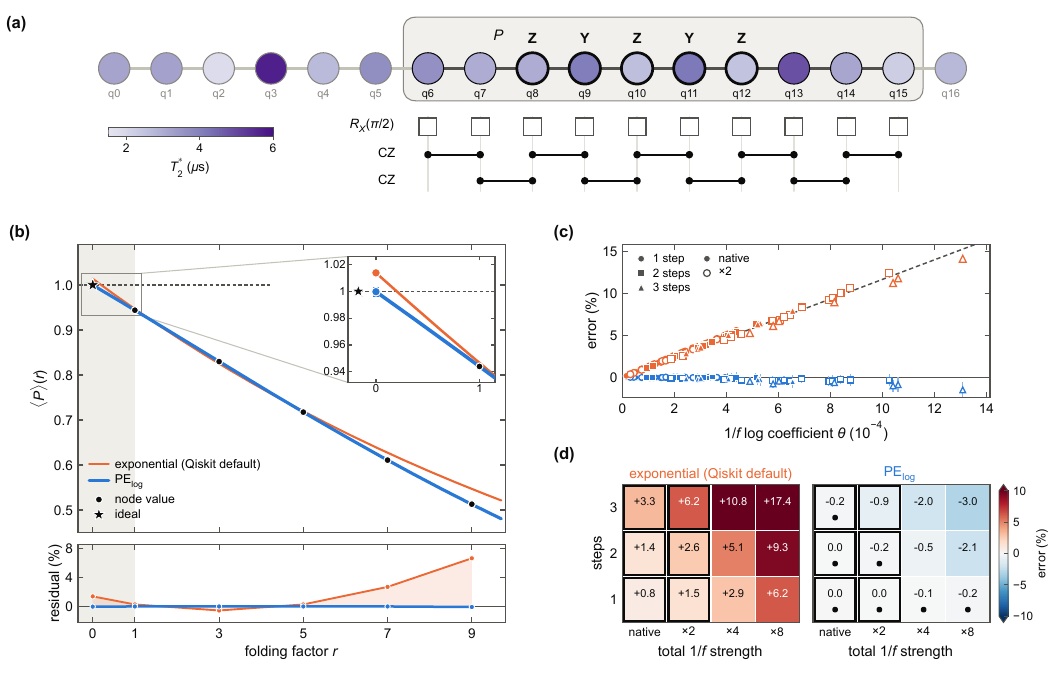}
\caption{\textbf{Multiqubit predictions link conventional bias to
the memory coefficient and show its suppression by PE\_log.}
All responses are model predictions, not processor measurements.
(a) Calibration-informed QZ01 chain and the q6--q15 segment used
in the model; the two-step target has support $ZYZYZ$ on q8--q12.
One kicked-Ising layer is shown; the virtual-$Z$ phases completing
the interactions are specified in SM, Sec.~S8.
(b) Two-step reference-noise response and inferred intercepts;
``native'' denotes the device-informed noise model. Exponential
extrapolation uses $(1,3,5)$ and PE\_log uses $(1,3,5,7)$.
Residual and shading conventions match Fig.~\ref{fig:correction}(a).
Node bars project sampling errors; intercept bars additionally
include numerical ensemble error in quadrature.
(c) All compatible nontrivial stabilizers in the selected measurement
bases: $31$, $15$ and $7$ at depths one, two and three, respectively.
Reference and doubled total $1/f$ strength give $106$ predictions
per estimator, plotted against the corresponding total $\theta$.
The dashed line is a pooled origin-constrained fit of the exponential
bias. Observables sharing a measurement basis have correlated
sampling errors; the bars show marginal projected standard deviations.
(d) Signed bias across depth and total $1/f$ strength. Boxes indicate settings selected for experimental investigation.
The corresponding hardware measurements are underway and will
be incorporated into a forthcoming revision of this manuscript. Dots denote bias smaller than the quadrature combination
of projected sampling and numerical ensemble errors. The error
budgets, model factorization and comparison to Richardson are
given in SM, Sec.~S8.}
\label{fig:application}
\end{figure*}

\paragraph*{Relation to existing methods.}
We remark that log-domain polynomial fitting is not new:
polyexponential models~\cite{GiurgicaTiron2020DigitalZNE} and the
cumulant formulation of PP-ZNE~\cite{Suchsland2026InformationDynamics}
already use it. The latter also connects single-exponential response
to delta-like circuit reactivity, including Clifford special cases,
and demonstrates persistent model bias without sampling noise.
Its Markovian channel description and the temporal-memory mechanism
here are complementary. The new coordinate is derived from noise
memory, with a spectrum--control coefficient, rather than introduced
as a generic extra polynomial term. It is absent from a finite
polynomial cumulant truncation; this is not a claim that a sufficiently
flexible analytic model cannot approximate a finite positive window.
Physical inverse engineering likewise precedes this work in
KIK~\cite{Henao2023AdaptiveKIK}. Our use of matching identifies which
offsets it cancels in the finite-band memory response.

\section{Numerical Illustrations}
\label{sec:numerical}
The numerical examples test the memory mechanism and its correction
without using hardware outcomes. All plotted responses are calculated
predictions; sampling error bars and RMSE curves are projections.
The single-qubit calculations use finite-duration QuTiP-based
propagators with symmetric phase kicks. The ten-qubit examples use
a calibration-informed model with separately calculated Markovian
and classical-noise factors. SM, Sec.~S8, specifies this approximation,
the smooth-band spectral regularization and the signal definitions;
these illustrations do not certify every bound for a sharp-cutoff
spectrum or an arbitrary hardware implementation.

\paragraph*{Single-qubit correction.}
The same response curve can support similar fits at noisy nodes yet
different zero-noise intercepts. Figure~\ref{fig:correction}(a)
demonstrates this for a nine-pulse sequence, denoted C8, and a
$1/f$ memory with inverse infrared rate $5\,\mu\mathrm{s}$.
The signal is the on/off contrast ratio
$y=(2P_{\rm on}-1)/(2P_{\rm off}-1)$, with $\chi=-\ln y$ and
$\chi(7)=0.45$. The calculated intercept biases are approximately
$+5.55\%$ for exponential extrapolation, $+1.71\%$ for
Richardson and $-0.26\%$ for PE\_log.
Both baselines use $(1,3,5)$; PE\_log uses $(1,3,5,7,9)$.
Thus the comparison tests these specified implementations, not
equal-parameter fits or a theorem about every node choice.
The projected RMSE curves in Fig.~\ref{fig:correction}(b)
separate persistent model bias from sampling uncertainty.
The parameter map in Fig.~\ref{fig:correction}(c) shows where
the leading four-coordinate model is accurate and where a
short memory window or stronger noise exposes the remainder.

\paragraph*{Multiqubit response.}
The forward coefficient also predicts how bias changes with the
observable. Figure~\ref{fig:application} considers one to three
Clifford kicked-Ising steps on a ten-qubit chain. All signed
stabilizer targets have ideal expectation one.
For independent local colored noise,
Eq.~\eqref{eq:fold-response} supplies
$\theta_{\rm total}=\sum_k A_k s_{2,k}/\pi$.
Across compatible stabilizers and the reference and doubled-strength
settings, the calculated exponential bias is approximately
proportional to this coefficient
[Fig.~\ref{fig:application}(c)]. The pooled through-origin slope
is $116.6$, with $R^2=0.997$; it is specific to this model and
fitting window, not a universal constant.
PE\_log suppresses the trend, although stronger noise and greater
depth leave visible residuals [Fig.~\ref{fig:application}(d)].
For the two-step reference target, the predicted biases are
$+1.38\%$ and $-0.03\%$ for exponential extrapolation and
four-node PE\_log, respectively.
The full target-observable comparison, including Richardson,
is given in SM, Sec.~S8.

\section{Discussion}
\label{sec:discussion}
Noise amplification reads temporal correlations as well as accumulating
local errors. The circuit-kernel relation makes that dependence
explicit, and exact block counting converts it into a response law
for full multiqubit folding. The $r^2\ln r$ coordinate and its
forward coefficient lead directly to a calculable Richardson bias
component and to response-matched cancellation. This gives a
physical reason to change the extrapolation basis, rather than
choosing its form solely by goodness of fit.

Three levels of the result should be distinguished. The structural
theory applies to arbitrary finite circuits with the declared
repeatable controls and noise covariance. The four-coordinate
estimator removes the derived logarithmic contribution, with
remainder and offset qualifications. Exact recovery of the ideal
value requires stronger matching or identifiable model information;
it is not implied by preserving the noiseless gate product.
The pulse-stretching comparison sharpens this distinction:
the shared logarithmic mechanism coexists with different physical
anchors.

Implementation of the estimator needs only folded circuits and the
same observable measured at the chosen nodes. Predicting $\theta$
or certifying the residual bound additionally requires the coupling
and ideal control response; logical gate labels alone do not supply
all that information. Complete forward and inverse blocks must
retain their prescribed physical schedules. State preparation and
measurement errors that are not amplified also require separate
treatment. The present $r$-dependent theorem concerns complete
global folding, not arbitrary local or two-qubit-only folding
schedules. The more general copy-count statement in the SM does
not extend the same logarithmic law to those schedules.

The approximation can fail when the sampled delays leave the
logarithmic memory window, when higher noise orders become large,
or when a small leading coefficient exposes control-shape terms.
A dominant non-Gaussian fluctuator requires higher correlators
beyond the covariance; stationarity and Gaussian remainder bounds
must not be assumed in that regime. Stronger noise or longer memory
does not automatically make the correction more accurate.
Statistical uncertainty is likewise not removed by a change of basis,
and a general finite-budget advantage is not claimed.

These limitations define direct tests rather than alternative fitting
rules. Controlled changes of the covariance can test the translated
memory relation in Eq.~\eqref{eq:curvature}; independently determined
spectral amplitudes and circuit responses can test the coefficient in
Eq.~\eqref{eq:fold-response}; extrapolation can then test removal of
the predicted bias without changing the chosen basis.
The theoretical conclusion is that memory supplies an identifiable
nonpolynomial response coordinate. Accounting for that coordinate
is distinct from, and necessary for assessing, the remaining
physical and statistical errors of ZNE.

Experimental tests of the memory-dependent folding response and
its correction by PE$_{\log}$ are underway on a superconducting
quantum processor. The present version establishes the theoretical
framework and numerical predictions. Experimental results will be
incorporated into an upcoming revision of this manuscript, providing
a unified theoretical and experimental investigation.

\clearpage
\onecolumngrid

% ---------- Supplemental Material (defined in sm.tex) ----------
% ============================================================================
% Supplemental Material for
%   Logarithmic Bias of Zero-Noise Extrapolation under $1/f$ Noise
%
% This file is \input by main.tex after the main bibliography.  It is not
% a standalone document: the preamble, the theorem environments and the
% reference list all live in main.tex.
% ============================================================================

\setcounter{section}{0}
\setcounter{equation}{0}
\setcounter{figure}{0}
\setcounter{table}{0}
\renewcommand{\thesection}{S\arabic{section}}
\renewcommand{\theequation}{S\arabic{equation}}
\renewcommand{\thefigure}{S\arabic{figure}}
\renewcommand{\thetable}{S\arabic{table}}

\begin{center}
{\large\bfseries SUPPLEMENTAL MATERIAL}\\[6pt]
{\large\bfseries Logarithmic Bias of Zero-Noise Extrapolation under $1/f$ Noise}\\[4pt]
Chenhui Wang, Weilong Wang, Haoran He, Ping Lv, and Zheng Shan\\[2pt]
{\small Laboratory for Advanced Computing and Intelligence Engineering, Zhengzhou 450001, Henan, China}
\end{center}

\allowdisplaybreaks

This supplement derives the response coordinates used for extrapolation in
the presence of time-local noise and stationary classical $1/f$ noise.
The two amplification protocols act differently on the time arguments of the
noise covariance. Their common logarithmic coordinate is therefore derived
separately, with the conditions on its coefficient and residual terms stated
for each protocol. The analysis retains first order in the white-noise
generator and second order in the colored-noise coupling. A statement about
this retained response is not an all-orders or finite-sample unbiasedness
claim.

\paragraph*{Notation.}
We set $\hbar=1$ and use natural logarithms and angular frequencies. The
superoperator norm is the trace-norm induced norm on the full complex matrix
space, unless stated otherwise. We distinguish the physical ensemble mean
$E$ from a sampled estimate $\widehat E$.
\begin{table}[htbp]
\centering
\caption{Notation and conventions used in this supplement.}
\label{tab:sm-notation}
\begin{tabular}{@{}p{0.23\textwidth}p{0.71\textwidth}@{}}
\toprule
Symbol & Definition \\
\midrule
$\rho_0$, $O$, $U$ & Input density operator, measured Hermitian observable,
and ideal target unitary. \\
$O_U$, $E^*$ & $U^\dagger OU$ and $\Tr(O_U\rho_0)$. \\
$T$, $c$, $r$ & Complete reference-block duration, pulse-stretch factor,
and complete-fold coordinate $r=2n+1$, respectively. \\
$n$ & Number of inverse--forward block pairs. \\
$\lambda$, $\mathcal L_w$ & Colored-noise coupling scale and time-local
white-noise generator, with white-noise rates included in $\mathcal L_w$. \\
$\mathsf A$, $C(\tau)$ & Spatial covariance matrix and scalar temporal
covariance, so $C_{\alpha\beta}(\tau)=\mathsf A_{\alpha\beta}C(\tau)$. \\
$\omega_{\rm IR}$, $\omega_{\rm UV}$ & Positive finite infrared and
ultraviolet angular-frequency cutoffs. \\
$\ell_T$ & $-\gamma_{\rm E}-\ln(\omega_{\rm IR}T)$. \\
$\Gamma_{0,\rm ps}$, $\Gamma_{0,\rm gf}$ & Protocol-specific
zero-frequency response coefficients. \\
$\theta_{\rm ps}$, $\theta_{\rm gf}$ & Logarithmic coefficients
$\lambda^2\Gamma_{0,\rm ps}/\pi$ and
$\lambda^2\Gamma_{0,\rm gf}/\pi$. \\
$\mathcal M_0$, $M_0$ & Fold zero-frequency moment superoperator and its
observable contraction; $M_0=8\Gamma_{0,\rm gf}$. \\
$\Omega$, $s_0$ & First-white-order native constant and constant coefficient
of the static colored response. Neither denotes the entire residual bias. \\
$\delta_{\rm gf}$, $R$ & Constant of the declared anchored folding
representation and its separately specified remainder. \\
$\Pi_F$, $\Pi_I$ & Forward and inverse Pauli-channel copy products
in the exact integer-node Clifford special case. \\
$\kappa_{\rm id}$ & Model-assisted identifiability margin. \\
$\mathcal B_A$, $h_{\rm bd}$ & Coupling-norm contraction and corresponding
observable error-bound scale, defined in Eq.~\eqref{eq:sm-fold-bound-scale}. \\
\bottomrule
\end{tabular}
\end{table}

\section{Response theory}
\label{sm:response-theory}

\subsection{Control and noise model}
Consider an arbitrary fixed finite-dimensional register. The Hilbert space
may include more than two levels per device; no product-state, commuting-gate,
or single-qubit restriction is imposed. Let $x$ denote the amplification
coordinate, $x=c$ for stretching or $x=r$ for complete folding, and let $D_x$
be the implemented duration. The ideal propagator satisfies
$U_x(0)=I$ and $U_x(D_x)=U$. Conditional on the classical noise realization,
the state evolves as
\begin{equation}
 \dot\rho_x(t)=-i\left[H_{0,x}(t)+\lambda\sum_\alpha
 \xi_\alpha(t)B_{x,\alpha}(t),\rho_x(t)\right]
 +\mathcal L_{w,x}(t)\rho_x(t).
 \label{sm:noise-model}
\end{equation}
The Hermitian coupling operators obey
$\|B_{x,\alpha}(t)\|_\infty\leq\beta_\alpha$ on the specified finite
node set. The white-noise generator is deterministic, bounded and time-local,
with GKSL form. Its rates are not rescaled together with the ideal control.
The classical fields are independent of the white-noise mechanism, centered,
jointly Gaussian and stationary. Stationarity and their covariance suffice
for the second-order response; Gaussianity is additionally used below to
bound all omitted colored orders.

For the finite-band $1/f$ sector we use
\begin{align}
 C_{\alpha\beta}(\tau)
 &=\E[\xi_\alpha(t+\tau)\xi_\beta(t)]
 =\frac{1}{2\pi}\int_{\mathbb R}
 S_{\alpha\beta}(\omega)e^{-i\omega\tau}\,d\omega,
 \label{sm:spectral-convention}\\
 S_{\alpha\beta}(\omega)
 &=\frac{\mathsf A_{\alpha\beta}}{|\omega|}
 \mathbf 1_{\{\omega_{\rm IR}<|\omega|<\omega_{\rm UV}\}},
 \qquad \mathsf A=\mathsf A^{\mathsf T}\succeq0,
 \quad 0<\omega_{\rm IR}<\omega_{\rm UV}<\infty.
 \label{sm:spectrum}
\end{align}
Thus the covariance has a common temporal shape and permits spatially
correlated noise. It does not include a frequency-dependent complex phase
in a cross-spectrum. The fields remain continuous in laboratory time across
successive blocks; folding does not reset the environment or replay the same
random trajectory. Finite-band spectra and the familiar logarithmic memory
associated with $1/f$ noise are reviewed in Ref.~\cite{Paladino2014OneOverF}.

\subsection{Ordered response and retained orders}
Define
\begin{equation}
 \widetilde\rho_x(t)=U_x^\dagger(t)\rho_x(t)U_x(t),\qquad
 b_{x,\alpha}(t)=U_x^\dagger(t)B_{x,\alpha}(t)U_x(t),\qquad
 \ad_B(X)=[B,X].
\end{equation}
Differentiation cancels the ideal-control commutator and gives
\begin{equation}
 \dot{\widetilde\rho}_x(t)=
 \left[\mathcal L_{w,x}^{I}(t)
 -i\lambda\sum_\alpha\xi_\alpha(t)\ad_{b_{x,\alpha}(t)}\right]
 \widetilde\rho_x(t),
 \label{sm:interaction-evolution}
\end{equation}
where $\mathcal L_{w,x}^{I}$ is the white-noise generator conjugated by the
ideal propagator. Iterating the integral equation yields its time-ordered
Dyson series. A single colored insertion averages to zero, whereas two
colored insertions contribute $(-i)^2=-1$. Retaining one white insertion and
two colored insertions therefore gives
\begin{align}
 \E\widetilde\rho_x(D_x)
 &=\rho_0+\mathcal W_x\rho_0-\lambda^2\mathcal K_x\rho_0
 +\mathcal R_{{\rm high},x}\rho_0,
 \label{sm:second-response}\\
 \mathcal W_x&=\int_0^{D_x}\mathcal L_{w,x}^{I}(t)\,dt,\\
 \mathcal K_x&=\sum_{\alpha\beta}\mathsf A_{\alpha\beta}
 \int_0^{D_x}dt\int_0^t ds\,
 C(t-s)\ad_{b_{x,\alpha}(t)}\ad_{b_{x,\beta}(s)}.
 \label{sm:ordered-response}
\end{align}
In particular,
\begin{equation}
 E(x)=E^*+\Tr(O_U\mathcal W_x\rho_0)
 -\lambda^2\Tr(O_U\mathcal K_x\rho_0)+R_{\rm high}(x).
 \label{sm:mean-response}
\end{equation}
The observable kernel is
\begin{equation}
 K_{\alpha\beta,x}(t,s)
 =\Tr\!\left(O_U[b_{x,\alpha}(t),[b_{x,\beta}(s),\rho_0]]\right).
 \label{sm:observable-kernel}
\end{equation}
Its contraction with the covariance is the complete second-order response.
For general controls and observables it cannot be replaced by a scalar,
nonnegative filter power without additional conditions: exchanging the two
times also changes the order of the two commutator superoperators.

A term with one white and one colored insertion also averages to zero.
Terms with one white and two colored insertions generally survive. Together
with second and higher white orders and fourth and higher colored orders,
they belong to $R_{\rm high}$. Accordingly, a quadratic regular coordinate
does not imply that all second-white-order terms have been retained. The
common comparison in this supplement uses the truncation in
Eq.~\eqref{sm:second-response}.

\subsection{A bound on omitted noise orders}
Set
\begin{equation}
 K_w(D)=\int_0^D\|\mathcal L_w^I(t)\|_{1\to1}\,dt,\qquad
 \varsigma=\sum_\alpha\beta_\alpha\sqrt{C_{\alpha\alpha}(0)},\qquad
 Z(D)=2\lambda^2D^2\varsigma^2.
 \label{sm:majorant-scales}
\end{equation}
Here $C_{\alpha\alpha}(0)$ includes the spatial spectral amplitude but not
the external factor $\lambda^2$.
\begin{proposition}[Mixed-noise remainder]
Under the bounded-control and Gaussian assumptions above,
\begin{equation}
 |R_{\rm high}(x)|\leq\|O\|_\infty
 \left[e^{K_w(D_x)+Z(D_x)}-1-K_w(D_x)-Z(D_x)\right].
 \label{sm:high-bound}
\end{equation}
\end{proposition}
\begin{proof}
The trace norm gives $\|\ad_B\|_{1\to1}\leq2\|B\|_\infty$.
Gaussian pairing and the covariance inequality
$|C_{\alpha\beta}(t-s)|\leq
\sqrt{C_{\alpha\alpha}(0)C_{\beta\beta}(0)}$ bound a sum of $2q$
colored insertions by
$(2q-1)!!(2|\lambda|\varsigma)^{2q}$. For $p$ white insertions,
summing over all interleavings of white and colored times gives
\begin{equation}
 \sum_{\substack{S\subset\{1,\ldots,p+2q\}\\|S|=p}}
 \int_{D>t_1>\cdots>t_{p+2q}>0}
 \prod_{j\in S}\|\mathcal L_w^I(t_j)\|_{1\to1}\,d^{p+2q}t
 =\frac{K_w(D)^pD^{2q}}{p!(2q)!}.
\end{equation}
This identity follows by integrating separately labeled white and colored
times over the full cube, partitioning it by time ordering, and dividing by
the $p!(2q)!$ permutations within the two classes. It is an identity for the
sum of interleavings, not for each interleaving separately. The complete
$(p,2q)$ class is consequently bounded by $K_w^p Z^q/(p!q!)$.

To justify averaging the full Dyson series, define
$\Xi_D=2|\lambda|\int_0^D\sum_\alpha\beta_\alpha|\xi_\alpha(t)|\,dt$.
For $\varsigma>0$, Jensen's inequality with channel weights
$\beta_\alpha\sqrt{C_{\alpha\alpha}(0)}/\varsigma$ and time weight $dt/D$
gives $\E e^{\Xi_D}\leq2e^{Z(D)}$; zero-variance channels can be omitted.
The bound follows from $e^{u|N|}\leq e^{uN}+e^{-uN}$ for a standard Gaussian
$N$. The pathwise absolute Dyson sum is bounded by $e^{K_w+\Xi_D}$,
whose expectation is finite. The averaged classes can therefore be summed
absolutely. Summing $K_w^pZ^q/(p!q!)$ and removing the retained classes
$(p,q)=(0,0),(1,0),(0,1)$ proves Eq.~\eqref{sm:high-bound}, after contracting
with the normalized input state and $O_U$.
\end{proof}
The right-hand side starts at second white order, fourth colored order and
mixed white--colored order. It is a bound, not a claim that those terms are
small for every duration or register size. The fixed finite nodes and the
chosen noise scales determine whether truncation is informative.

\section{Amplification and time arguments}
\label{sm:time-transformations}

\subsection{Pulse stretching}
Ideal pulse stretching rescales the complete deterministic Hamiltonian,
\begin{equation}
 H_{0,c}(t)=\frac1c H_0(t/c),\qquad 0\leq t\leq cT.
 \label{sm:stretch-hamiltonian}
\end{equation}
It includes retained drifts, entangling interactions and idle durations;
rescaling only the nominal duration of a logical gate is not this
Hamiltonian identity. A sufficient coupling condition is
$B_{c,\alpha}(cu)=B_\alpha(u)$, automatic for a fixed laboratory-frame
coupling. For the white sector we additionally require
$\mathcal L_{w,c}^{I}(cu)=\mathcal L_w^I(u)$ with laboratory rates fixed.
These are response-covariance conditions for the idealized stretching model,
not assumptions about access to a particular hardware interface.
Pulse-based noise scaling was introduced and implemented in
Refs.~\cite{Temme2017ErrorMitigation,Kandala2019ComputationalReach}; its
spectral consequences under time-correlated noise are discussed in
Ref.~\cite{Schultz2022TimeCorrelatedZNE}.

Let $V_c(u)=U_c(cu)$. It solves
$i\dot V_c(u)=cH_{0,c}(cu)V_c(u)=H_0(u)V_c(u)$ with $V_c(0)=I$.
Uniqueness of this initial-value problem gives
\begin{equation}
 U_c(cu)=U_0(u),\qquad b_{c,\alpha}(cu)=b_\alpha(u),\qquad
 U_c(cT)=U.
 \label{sm:stretch-time}
\end{equation}
No interchange of noncommuting Hamiltonian terms is involved. The ideal
trajectory, pulled-back observable and observable kernel are unchanged in
circuit time. Substitution of $t=cu$, $s=cv$ in the ordered response gives
\begin{equation}
 \Delta_{\rm ps}^{(2)}(c)
 =-\lambda^2c^2\sum_{\alpha\beta}\mathsf A_{\alpha\beta}
 \int_0^Tdu\int_0^u dv\,
 C\bigl(c(u-v)\bigr)K_{\alpha\beta}(u,v).
 \label{sm:stretch-kernel}
\end{equation}
Stretching thus changes the time separation read from a stationary
environment, while keeping the ideal-control kernel fixed.

For later use define the triangular frequency response
\begin{equation}
 G_{\alpha\beta}(\omega)=\int_0^Tdu\int_0^u dv\,
 e^{-i\omega(u-v)}K_{\alpha\beta}(u,v).
 \label{sm:frequency-kernel}
\end{equation}
The same substitution proves
$G_{\alpha\beta,c}(\omega)=c^2G_{\alpha\beta}(c\omega)$.
Fourier transformation therefore reads the laboratory spectrum at
rescaled frequency $\omega/c$. The white term, by contrast, is
\begin{equation}
 \Tr(O_U\mathcal W_c\rho_0)=c\,a_{1,\rm ps},\qquad
 a_{1,\rm ps}=\int_0^T\Tr\!\left(O_U\mathcal L_w^I(u)\rho_0\right)du.
 \label{sm:stretch-white}
\end{equation}
It remains present when $1/f$ noise is also present. If the entire second
white order is retained, its double time integral gives an additional
regular $c^2$ term under the same covariance condition. That is a different
truncation from Eq.~\eqref{sm:second-response}; mixed terms must still be
accounted for.

\subsection{Complete-circuit folding}
\label{sec:sm-fold-count}

Let $F$ implement the complete target circuit $U$, and let $I$ implement
$U^\dagger$. The two blocks have the same duration $T$, although the gates
within either block need not have equal durations. Complete global folding
runs the chronological sequence
\begin{equation}
 F,I,F,I,\ldots,I,F,\qquad r=2n+1,\qquad n=0,1,\ldots .
 \label{eq:sm-fold-sequence}
\end{equation}
There are $n+1$ forward blocks and $n$ inverse blocks. Their ideal endpoint
is $U(U^\dagger U)^n=U$, and their total duration is $D=rT$.
For a circuit $U=G_J\cdots G_1$, the inverse block executes
$G_J^\dagger,\ldots,G_1^\dagger$ in chronological order. Thus the construction
includes both single-qubit and entangling gates; it is not a repetition of
the CZ gates alone.
This is complete global unitary folding in the standard digital-ZNE
construction~\cite{GiurgicaTiron2020DigitalZNE}.

Each occurrence of a given block repeats the same physical implementation,
including idle intervals, parallel layers and coupling sensitivities. The
compiler preserves the inserted inverse--forward pairs, and the physical
noise parameters do not change with the selected fold node. The laboratory
white-noise generators repeat as fixed, bounded profiles
$\mathcal L^{\rm lab}_{w,F}(u)$ and $\mathcal L^{\rm lab}_{w,I}(u)$,
$0\leq u<T$. These profiles may differ from one another. Repeatability refers
to their average generators, not to identical random errors in every copy.
The colored process remains continuous across block boundaries and is not
reset when an inverse block begins.

The repeated colored-coupling profiles must be compared in a common
interaction frame. With $U_F(T)=U$ and $U_I(T)=U^\dagger$, they are
\begin{align}
 f_\alpha(u)&=U_F^\dagger(u)B_{F,\alpha}(u)U_F(u),\\
 i_\alpha(u)&=U^\dagger U_I^\dagger(u)B_{I,\alpha}(u)U_I(u)U.
 \label{eq:sm-common-profiles}
\end{align}
The outer $U^\dagger$ and $U$ in the second expression account for the
ideal propagation preceding an inverse block. Write
$b_{F,\alpha}=f_\alpha$ and $b_{I,\alpha}=i_\alpha$.
For white noise, with $\mathcal C_V(X)=VXV^\dagger$, the analogous profiles are
\begin{align}
 \mathcal L^I_{w,F}(u)
 &=\mathcal C_{U_F(u)}^{-1}\mathcal L^{\rm lab}_{w,F}(u)
   \mathcal C_{U_F(u)},\\
 \mathcal L^I_{w,I}(u)
 &=\mathcal C_U^{-1}\mathcal C_{U_I(u)}^{-1}
   \mathcal L^{\rm lab}_{w,I}(u)\mathcal C_{U_I(u)}\mathcal C_U.
\end{align}
Define $\mathcal W_X=\int_0^T\mathcal L^I_{w,X}(u)\,du$.
Their first-order contribution is exactly affine in the fold coordinate:
\begin{equation}
 \mathcal W_r=(n+1)\mathcal W_F+n\mathcal W_I
 =r\mathcal W_++\mathcal W_-,\qquad
 \mathcal W_\pm=\frac{\mathcal W_F\pm\mathcal W_I}{2}.
 \label{eq:sm-white-affine}
\end{equation}
Consequently, its observable contribution is $a_w r+\Omega$, where
$a_w=\operatorname{Tr}(O_U\mathcal W_+\rho_0)$ and
$\Omega=\operatorname{Tr}(O_U\mathcal W_-\rho_0)$.
The ideal inverse relation alone does not set $\Omega$ to zero.

More generally, multiply the white generator by a formal strength
$\varepsilon$ and let $\mathcal A(\varepsilon)$ and
$\mathcal B(\varepsilon)$ be its fixed forward and inverse block maps in
the common frame. Both equal the identity at zero strength. The folded
map is $\mathcal A(\mathcal B\mathcal A)^n$. Near zero, the convergent
logarithm $\mathcal V=\log(\mathcal B\mathcal A)$ has no constant term,
so the coefficient of $\varepsilon^p$ in $e^{n\mathcal V}$ is a polynomial
in $n$ of degree at most $p$. The same is true after multiplication by
$\mathcal A$ and conversion to $r=2n+1$. Thus repeatable time-local noise
has regular polynomial coordinates order by order, even without commuting
forward and inverse generators. This argument allows constant terms; it
does not remove them. The explicit remainder certificate here retains
only first white order.

For the colored second-order response, define the within-block and
cross-block superoperators
\begin{align}
 \mathcal R_X&=\sum_{\alpha\beta}\mathsf A_{\alpha\beta}
 \int_0^Tdu\int_0^u dv\, C(u-v)
 \operatorname{ad}_{b_{X,\alpha}(u)}
 \operatorname{ad}_{b_{X,\beta}(v)},\\
 \mathcal K_{XY}(d)&=\sum_{\alpha\beta}\mathsf A_{\alpha\beta}
 \int_0^Tdu\int_0^Tdv\, C(dT+u-v)
 \operatorname{ad}_{b_{X,\alpha}(u)}
 \operatorname{ad}_{b_{Y,\beta}(v)}.
 \label{eq:sm-crossblock}
\end{align}
Here $X$ labels the later block, $Y$ the earlier block, and $d$ is their
integer start-time separation in units of $T$. The superoperator ordering
is inherited from the time-ordered response and has not been symmetrized.
Partitioning its integration triangle by blocks gives
\begin{align}
 \mathcal K_n={}&(n+1)\mathcal R_F+n\mathcal R_I\notag\\
 &+\sum_{k=1}^{n}\bigl[(n+1-k)\mathcal K_{FF}(2k)
                     +(n-k)\mathcal K_{II}(2k)\bigr]\notag\\
 &+\sum_{k=0}^{n-1}(n-k)
     \bigl[\mathcal K_{IF}(2k+1)+\mathcal K_{FI}(2k+1)\bigr].
 \label{eq:sm-exact-count}
\end{align}
To see that the counting is exhaustive, within-block pairs occur in each
of the $n+1$ forward and $n$ inverse copies. Equal-type blocks have even
separation $2k$, with $n+1-k$ forward pairs and $n-k$ inverse pairs.
Opposite-type blocks have odd separation $2k+1$, with $n-k$ pairs in each
time ordering. Every ordered time pair belongs to exactly one of these
sets. Equation~\eqref{eq:sm-exact-count} therefore requires no mirror
relation between the two implementations.

\section{Curvature and noise memory}
\label{sm:curvature-memory}
\subsection{Second differences isolate translated memory}
\label{sec:sm-curvature-fold}

For the odd nodes $r=2n+1$, define the unnormalized second difference
\begin{equation}
 \Delta_2^2 E(r)=E(r+2)-2E(r)+E(r-2),\qquad n\geq1.
 \label{eq:sm-difference-definition}
\end{equation}
There is no division by the squared node spacing. In
Eq.~\eqref{eq:sm-exact-count}, every separation already present at the
three adjacent values of $n$ has an affine multiplicity and cancels.
The within-block terms cancel for the same reason. Only the newly exposed
cross-block separations remain:
\begin{equation}
 \Delta_n^2\mathcal K_n
 =\mathcal K_{FF}(r+1)+\mathcal K_{II}(r-1)
  +\mathcal K_{IF}(r)+\mathcal K_{FI}(r),
 \label{eq:sm-count-difference}
\end{equation}
where $\Delta_n^2\mathcal K_n=\mathcal K_{n+1}-2\mathcal K_n+
\mathcal K_{n-1}$.

Introduce two auxiliary profiles of duration $2T$:
$P=(I,F)$ and $Q=(F,I)$, with the notation specifying which common-frame
profile occupies each half. They are integration devices, not additional
circuits to be executed. Splitting the square $[0,2T]^2$ into its four
quadrants rewrites Eq.~\eqref{eq:sm-count-difference} as
\begin{equation}
 \Delta_n^2\mathcal K_n
 =\sum_{\alpha\beta}\mathsf A_{\alpha\beta}
 \int_0^{2T}du\int_0^{2T}dv\,
 C(rT+u-v)\operatorname{ad}_{P_\alpha(u)}
                  \operatorname{ad}_{Q_\beta(v)}.
 \label{eq:sm-memory-rectangle}
\end{equation}
For example, the later half of $P$ and the earlier half of $Q$ yield
$C((r+1)T+u'-v')$ and hence the $FF$ term. The opposite quadrant gives
the $II$ separation $(r-1)T$; the other two give $rT$. Since $r\geq3$,
all separations are positive and the rectangle avoids the time diagonal.

Equation~\eqref{eq:sm-memory-rectangle} is the precise curvature--memory
relation: the second difference reads a control-weighted range of memory
values around $rT$. Replacing the integrand by a single value $C(rT)$ is
an additional approximation, not this identity. The first-white-order
affine contribution in Eq.~\eqref{eq:sm-white-affine} is removed by the
same second difference; omitted mixed and higher orders must still be
retained in the response error account.

Define integrated block responses
\begin{equation}
 Q_{F,\alpha}=\int_0^T f_\alpha(u)\,du,\qquad
 Q_{I,\alpha}=\int_0^T i_\alpha(u)\,du,\qquad
 Q_{+,\alpha}=\frac{Q_{F,\alpha}+Q_{I,\alpha}}{2}.
\end{equation}
Both auxiliary profiles integrate to $2Q_{+,\alpha}$. The zeroth
rectangle moment is therefore
\begin{equation}
 \mathcal M_0=4\sum_{\alpha\beta}\mathsf A_{\alpha\beta}
 \operatorname{ad}_{Q_{+,\alpha}}\operatorname{ad}_{Q_{+,\beta}},
 \qquad M_0=\operatorname{Tr}(O_U\mathcal M_0\rho_0).
 \label{eq:sm-memory-weight}
\end{equation}
This circuit weight follows from the complete fold counting. It is not
an assumed low-frequency filter response.

\subsection{Memory classes}
Four illustrative memory classes differ in the
function integrated against the same response kernel. White noise is
represented by the time-local generator: its first-order fold response
is affine and has zero second difference. For a frozen classical field,
$C(\tau)=C_0$ throughout the circuit, the second-order fold response is a
quadratic polynomial, including the possible constant $s_0$ derived in
Sec.~\ref{sm:infrared-constant}. Under the stated matching and commuting-
observable conditions that constant vanishes, so a quadratic-annihilating
linear extrapolation has zero bias from this frozen second-order sector.
This does not assert zero total bias at finite noise strength.

An Ornstein--Uhlenbeck process has a decaying covariance of the form
$C(\tau)=C_0 e^{-|\tau|/\tau_c}$. The exact rectangle identity still
applies: it integrates this decay over the block-dependent separations,
rather than automatically returning $C(rT)$. In the $1/f$ case, the
finite-band covariance has an intermediate-time logarithmic form. Its
discrete integration produces the logarithmic response coordinate derived
next. A memory diagnostic must use the complete response integral.
A pointwise memory curve is an interpretation of the curvature, not a
substitute for that integral when memory changes appreciably within one
block.

\section{Finite-band logarithmic response}
\label{sm:logarithmic-response}

\subsection{Pulse-stretching logarithm}
Define the even real contracted response
\begin{equation}
 F_O(\omega)=\operatorname{Re}\sum_{\alpha\beta}
 \mathsf A_{\alpha\beta}G_{\alpha\beta}(\omega),\qquad
 \Gamma_{0,\rm ps}=F_O(0).
 \label{sm:stretch-plateau}
\end{equation}
The finite-band second-order contribution is exactly
\begin{equation}
 \Delta_{\rm ps}^{(2)}(c)
 =-\frac{\lambda^2c^2}{\pi}
 \int_{c\omega_{\rm IR}}^{c\omega_{\rm UV}}
 \frac{F_O(\omega)}{\omega}\,d\omega.
 \label{sm:stretch-band-integral}
\end{equation}
The prefactor $c^2$ follows from time rescaling and the $1/|\omega|$
spectrum. The logarithm follows from the finite-duration response at low
frequency; it is not an assumed regression term.

The reality of the time-domain kernel and $|\cos u-1|\leq u^2/2$ give
\begin{align}
 |F_O(\omega)-\Gamma_{0,\rm ps}|&\leq L_2\omega^2,\\
 L_2&=\frac12\sum_{\alpha\beta}|\mathsf A_{\alpha\beta}|
 \int_0^Tdu\int_0^u dv\,(u-v)^2|K_{\alpha\beta}(u,v)|.
 \label{sm:stretch-low-frequency-bound}
\end{align}
To separate a constant from the high-frequency response, suppose additionally
that fixed $\omega_R>0$, $\beta>0$ and $L_\infty<\infty$ satisfy
\begin{equation}
 |F_O(\omega)|\leq L_\infty(\omega_R/|\omega|)^\beta,
 \qquad |\omega|\geq\omega_R.
 \label{sm:stretch-tail-assumption}
\end{equation}
Without this condition the finite-band integral remains the applicable
expression. For the fitted window assume
\begin{equation}
 c_{\max}\omega_{\rm IR}<\omega_R<c_{\min}\omega_{\rm UV}.
 \label{sm:stretch-window}
\end{equation}
Set
\begin{align}
 J_O(\omega_R)&=\int_0^{\omega_R}
 \frac{F_O(\omega)-\Gamma_{0,\rm ps}}{\omega}\,d\omega
 +\int_{\omega_R}^{\infty}\frac{F_O(\omega)}{\omega}\,d\omega,\\
 e_{\rm edge}(c)&=-\int_0^{c\omega_{\rm IR}}
 \frac{F_O(\omega)-\Gamma_{0,\rm ps}}{\omega}\,d\omega
 -\int_{c\omega_{\rm UV}}^{\infty}\frac{F_O(\omega)}{\omega}\,d\omega.
\end{align}
Splitting the integral at $\omega_R$ and adding and subtracting its
zero-frequency value yields the identity
\begin{equation}
 \int_{c\omega_{\rm IR}}^{c\omega_{\rm UV}}
 \frac{F_O(\omega)}{\omega}\,d\omega
 =\Gamma_{0,\rm ps}\ln\frac{\omega_R}{c\omega_{\rm IR}}
 +J_O(\omega_R)+e_{\rm edge}(c).
\end{equation}
Consequently,
\begin{align}
 E_{\rm ps}(c)&=E^*+a_{1,\rm ps}c+a_{2,\rm ps}c^2
 +\theta_{\rm ps}c^2\ln c+R_{\rm ps}(c),
 \label{sm:stretch-log}\\
 \theta_{\rm ps}&=\frac{\lambda^2\Gamma_{0,\rm ps}}{\pi},\qquad
 a_{2,\rm ps}=-\frac{\lambda^2}{\pi}
 \left[\Gamma_{0,\rm ps}\ln\frac{\omega_R}{\omega_{\rm IR}}
 +J_O(\omega_R)\right],\\
 R_{\rm ps}(c)&=-\frac{\lambda^2c^2}{\pi}e_{\rm edge}(c)
 +R_{\rm high}(c),\\
 |e_{\rm edge}(c)|&\leq\frac{L_2}{2}(c\omega_{\rm IR})^2
 +\frac{L_\infty}{\beta}
 \left(\frac{\omega_R}{c\omega_{\rm UV}}\right)^\beta.
 \label{sm:stretch-edge}
\end{align}
The logarithmic coefficient includes the control, state, observable and
spatial covariance through $\Gamma_{0,\rm ps}$; it is not determined by the
noise spectrum alone. If this overlap vanishes, the displayed logarithmic
coefficient vanishes. The edge correction remains of colored-noise second
order and is not part of the fourth-order Gaussian remainder. An ultraviolet
bound independent of $c$ does not imply that the corresponding physical
contribution has been removed by extrapolation.

\subsection{The zero-duration anchor}
The finite-window logarithmic coordinate must be distinguished from the exact
fixed-cutoff response as $c\to0$. Returning to the laboratory frequency gives
\begin{align}
 \Delta_{\rm ps}^{(2)}(c)
 &=-\frac{\lambda^2c^2}{\pi}
 \int_{\omega_{\rm IR}}^{\omega_{\rm UV}}
 \frac{F_O(c\omega)}{\omega}\,d\omega\\
 &=-\frac{\lambda^2\Gamma_{0,\rm ps}}{\pi}
 c^2\ln\frac{\omega_{\rm UV}}{\omega_{\rm IR}}+\varepsilon_0(c),\qquad
 |\varepsilon_0(c)|\leq\frac{\lambda^2L_2}{2\pi}
 c^4(\omega_{\rm UV}^2-\omega_{\rm IR}^2).
 \label{sm:stretch-zero-expansion}
\end{align}
The exact edge term cancels the apparent logarithmic singularity when
Eq.~\eqref{sm:stretch-window} no longer holds. A finite-band system is not
claimed to have a singular physical response at zero duration.

There is also a nonperturbative anchor for the ideally rescaled family.
Extend the stated control and coupling covariance mathematically to $c>0$,
without assuming that hardware implements $c<1$, and write
$\mathcal L_w^I(u)=\gamma\widetilde{\mathcal D}(u)$ and
$k_D=\int_0^T\|\widetilde{\mathcal D}(u)\|_{1\to1}du$.
Variation of constants relative to $\rho_0$, followed by trace-norm
contraction of the conditional GKSL propagator on Hermitian operators, gives
\begin{equation}
 |E_{\rm ps}(c)-E^*|
 \leq\|O\|_\infty c\left[\gamma k_D
 +2|\lambda|T\sum_\alpha\beta_\alpha
 \sqrt{C_{\alpha\alpha}(0)}\right]\longrightarrow0.
 \label{sm:stretch-anchor}
\end{equation}
The stochastic source is bounded using
$\|[B,\rho_0]\|_1\leq2\|B\|_\infty$ and
$\E|\xi_\alpha|\leq\sqrt{C_{\alpha\alpha}(0)}$ on the interval of
length $cT$. This proves the ideal endpoint of the rescaled family, not
unbiasedness of a finite-node fit. Its weighted edge and higher-order terms
remain in the estimator error.

\subsection{Finite-band memory and its logarithmic window}
\label{sm:finite-band}
\label{sec:sm-fold-band}

For the spectral convention in Eq.~\eqref{sm:spectral-convention}, the normalized finite-band
covariance is
\begin{equation}
 C(\tau)=\frac{\operatorname{Ci}(\omega_{\rm UV}|\tau|)
                  -\operatorname{Ci}(\omega_{\rm IR}|\tau|)}{\pi},
 \qquad \tau\ne0.
 \label{eq:sm-band-ci}
\end{equation}
At the origin its finite value follows from the original spectral
integral. For $x>0$,
\begin{equation}
 \operatorname{Ci}(x)=\gamma_{\rm E}+\ln x+
 \int_0^x\frac{\cos u-1}{u}\,du,
 \qquad
 |\operatorname{Ci}(x)-\gamma_{\rm E}-\ln x|\leq\frac{x^2}{4}.
 \label{eq:sm-ci-small}
\end{equation}
The inequality follows from $|1-\cos u|\leq u^2/2$.
Integration by parts also gives $|\operatorname{Ci}(x)|\leq2/x$.
With $\ell_T=-\gamma_{\rm E}-\ln(\omega_{\rm IR}T)$, the corresponding
auxiliary logarithmic kernel is
\begin{equation}
 C_{\log}(\tau)=\frac{\ell_T-\ln(|\tau|/T)}{\pi}.
 \label{eq:sm-log-memory}
\end{equation}
Its singularity at zero is integrable. It organizes the finite-window
response, but is not a physical covariance replacement at all times.

The ultraviolet bound $2/(\omega_{\rm UV}|\tau|)$ cannot be integrated
directly across the time diagonal. Instead, using
Eq.~\eqref{eq:sm-ci-small} below unit argument and $2/x$ above it yields,
for $\omega_{\rm UV}D\geq1$,
\begin{equation}
 \int_0^D(D-t)|\operatorname{Ci}(\omega_{\rm UV}t)|\,dt
 \leq\frac{2D}{\omega_{\rm UV}}
       [1+\ln(\omega_{\rm UV}D)].
\end{equation}
Let $\|B_\alpha(t)\|_\infty\leq\beta_\alpha$ and set
\begin{equation}
 \mathcal B_A=\sum_{\alpha\beta}|\mathsf A_{\alpha\beta}|
                           \beta_\alpha\beta_\beta,\qquad
 h_{\rm bd}=\frac{\lambda^2\|O\|_\infty\mathcal B_A T^2}{\pi}.
 \label{eq:sm-fold-bound-scale}
\end{equation}
Using $\|\operatorname{ad}_X\operatorname{ad}_Y\|_{1\to1}
\leq4\|X\|_\infty\|Y\|_\infty$, the change in the colored second-order
observable caused by replacing $C$ with $C_{\log}$ is bounded by
\begin{equation}
 \epsilon_{\rm edge}(r)\leq
 \frac{\lambda^2\|O\|_\infty\mathcal B_A}{\pi}
 \left\{\frac{\omega_{\rm IR}^2D^4}{12}
 +\frac{8D}{\omega_{\rm UV}}
      [1+\ln(\omega_{\rm UV}D)]\right\},\qquad D=rT.
 \label{eq:sm-fold-edge-bound}
\end{equation}
Indeed, an ordered integral of a nonnegative function of $t-s$ equals
$\int_0^D(D-u)g(u)\,du$; the infrared part then uses
$\int_0^D(D-u)u^2du=D^4/12$, and the ultraviolet part uses the integral
bound above. The cutoffs are fixed throughout this comparison. The
resulting error must be assessed on the actual node window; the existence
of a $1/f$ component alone does not make it negligible.

\subsection{The discrete logarithmic hierarchy}
\label{sec:sm-fold-hierarchy}

Let $y_n=-\lambda^2\operatorname{Tr}(O_U\mathcal K_n^{\log}\rho_0)$
denote only the colored second-order response computed with $C_{\log}$.
It excludes both $E^*$ and the white-noise contribution. For $r=2n+1\geq3$,
$|u-v|/(rT)\leq2/3$, so the expansion
$-\ln(1+z)=\sum_{p\geq1}(-1)^p z^p/p$ converges uniformly and absolutely
on the rectangle. Define its time moments by
\begin{equation}
 \mathcal M_p=\sum_{\alpha\beta}\mathsf A_{\alpha\beta}
 \int_0^{2T}du\int_0^{2T}dv\,(u-v)^p
 \operatorname{ad}_{P_\alpha(u)}\operatorname{ad}_{Q_\beta(v)}.
\end{equation}
Termwise integration of Eq.~\eqref{eq:sm-memory-rectangle} gives
\begin{align}
 \Delta_n^2y_n&=d_0+d_L\ln r+\sum_{p=1}^{\infty}\frac{d_p}{r^p},\\
 d_L&=\frac{\lambda^2}{\pi}\operatorname{Tr}(O_U\mathcal M_0\rho_0),
 \qquad d_0=-\ell_Td_L,\\
 d_p&=-\frac{\lambda^2(-1)^p}{\pi pT^p}
        \operatorname{Tr}(O_U\mathcal M_p\rho_0).
 \label{eq:sm-fold-difference-series}
\end{align}
All these coefficients belong to colored-noise second order. The inverse
powers of $r$ describe control-shape corrections, not higher powers of
the noise strength.

Truncating the series after $p=P$ leaves a second-difference remainder
$e_P(r)$ with the explicit bound
\begin{equation}
 |e_P(r)|\leq32h_{\rm bd}\sum_{p=P+1}^{\infty}
              \frac{(2/r)^p}{p(p+1)(p+2)},\qquad r\geq3.
 \label{eq:sm-moment-tail}
\end{equation}
This follows by combining the double-commutator norm bound with
\begin{equation}
 \int_0^{2T}du\int_0^{2T}dv\,|u-v|^p
 =\frac{2^{p+3}T^{p+2}}{(p+1)(p+2)}.
\end{equation}
The double discrete integral is expressed by
\begin{equation}
 \Phi_{\log}(n)=\sum_{k=1}^{n-1}(n-k)\ln(2k+1),\qquad
 \Phi_p(n)=\sum_{k=1}^{n-1}\frac{n-k}{(2k+1)^p},
\end{equation}
with both sums zero at $n=0,1$. The complete anchored expansion at any
finite $n$ is
\begin{align}
 y_n={}&y_0+n(y_1-y_0)+\frac{d_0}{2}n(n-1)
 +d_L\Phi_{\log}(n)+\sum_{p=1}^{P}d_p\Phi_p(n)\notag\\
 &+\sum_{k=1}^{n-1}(n-k)e_P(2k+1).
 \label{eq:sm-discrete-hierarchy}
\end{align}
For a sequence $g_k$, the sum $S_n=\sum_{k=1}^{n-1}(n-k)g_k$ obeys
$\Delta_n^2S_n=g_n$. Hence the right-hand side has the required second
difference and agrees with $y_n$ at $n=0,1$. Their difference is affine
and vanishes at those two points, proving the identity.

The hierarchy therefore contains more than one nonpolynomial coordinate.
Its four-function reduction below controls the omitted shape terms
explicitly instead of treating them as zero. Nor may a decaying
second-difference tail be absorbed without tracking its constant:
if $|\rho_k|\leq K(2k+1)^{-q}$ with $q>2$, then
\begin{equation}
 \sum_{k=1}^{n-1}(n-k)\rho_k
 =nA_\rho-B_\rho+\sum_{k=n+1}^{\infty}(k-n)\rho_k,
 \quad A_\rho=\sum_{k\geq1}\rho_k,\quad
 B_\rho=\sum_{k\geq1}k\rho_k.
\end{equation}
In terms of $r$, the affine part is
$A_\rho r/2-A_\rho/2-B_\rho$. Removing the decaying tail from an
asymptotic expression does not remove this constant contribution.

\subsection{A four-function representation with a retained remainder}
\label{sec:sm-fold-four-functions}

The leading logarithmic coefficient is fixed by the zeroth response moment:
\begin{equation}
 \theta_{\rm gf}=\frac{d_L}{8}
 =\frac{\lambda^2}{2\pi}\operatorname{Tr}\!\left[
 O_U\sum_{\alpha\beta}\mathsf A_{\alpha\beta}
 [Q_{+,\alpha},[Q_{+,\beta},\rho_0]]\right],
 \qquad |\theta_{\rm gf}|\leq2h_{\rm bd}.
 \label{eq:sm-fold-theta}
\end{equation}
Thus $\Gamma_{0,\rm gf}$, defined by
$\theta_{\rm gf}=\lambda^2\Gamma_{0,\rm gf}/\pi$, obeys
$M_0=8\Gamma_{0,\rm gf}$. No assumption of a nonzero response is needed
for the structural representation.

To establish the reduction on finite odd nodes, set $f(r)=r^2\ln r$.
Twice applying the fundamental theorem of calculus gives
$\Delta_2^2f(r)=\int_{-2}^{2}(2-|u|)f''(r+u)\,du$.
Since $f''(r)=2\ln r+3$,
\begin{equation}
 |\Delta_2^2f(r)-8\ln r-12|
 \leq\frac{8}{3r^2(1-4/r^2)}\leq\frac{24}{5r^2},\qquad r\geq3.
 \label{eq:sm-r2log-difference-bound}
\end{equation}
The bound follows by writing the difference as
$2\int_0^2(2-u)\ln(1-u^2/r^2)\,du$ and applying
$-\ln(1-x)\leq x/(1-x)$.
Meanwhile, Eq.~\eqref{eq:sm-moment-tail} at $P=0$ implies
$|\Delta_n^2y_n-d_0-d_L\ln r|\leq16h_{\rm bd}/r$.
Define
\begin{equation}
 a_{2,\rm gf}=-\theta_{\rm gf}(\ell_T+3/2),\qquad
 z_n=y_n-\theta_{\rm gf}r_n^2\ln r_n-a_{2,\rm gf}r_n^2,
 \qquad B_{\rm sh}=\frac{96}{5}h_{\rm bd}.
\end{equation}
The constant and logarithmic second-difference terms cancel, leaving
$|\Delta_n^2z_n|\leq B_{\rm sh}/r_n$.

Fix the mathematical anchor at $n=0,1$, before selecting any estimator:
\begin{equation}
 b_1=\frac{z_1-z_0}{2},\qquad
 \delta_{\rm col}=z_0-b_1
 =\frac{3y_0-y_1}{2}+\frac92\theta_{\rm gf}\ln3+3a_{2,\rm gf}.
 \label{eq:sm-fold-anchor}
\end{equation}
These quantities define a representation; they do not require additional
calibration experiments. Discrete integration gives the exact logarithmic-
kernel identity
\begin{equation}
 y_n=\delta_{\rm col}+b_1r_n+a_{2,\rm gf}r_n^2
     +\theta_{\rm gf}r_n^2\ln r_n+R_{\rm sh}(n),\qquad
 R_{\rm sh}(n)=\sum_{j=1}^{n-1}(n-j)\Delta_j^2z_j.
 \label{eq:sm-fold-anchored-four}
\end{equation}
Here $R_{\rm sh}(0)=R_{\rm sh}(1)=0$, and
\begin{equation}
 |R_{\rm sh}(n)|\leq\frac{B_{\rm sh}}{2}n(1+\ln n),\qquad n\geq1.
 \label{eq:sm-fold-shape-bound}
\end{equation}
The remainder is still of colored-noise second order. Its lower growth
in $r$ is not a higher-noise-order approximation.

For $\ell_T\geq\ln3$, the auxiliary logarithmic covariance is nonnegative
throughout the $r=1,3$ anchor triangles. Integrating its absolute response
and applying Eq.~\eqref{eq:sm-fold-anchor} yields
\begin{equation}
 |\delta_{\rm col}|\leq18h_{\rm bd}(\ell_T+3/2).
 \label{eq:sm-fold-color-offset-bound}
\end{equation}
In detail, $|y_n|\leq2h_{\rm bd}r^2(\ell_T-\ln r+3/2)$ at the two
anchors, obtained from
$\int_0^D(D-t)[\ell_T-\ln(t/T)]dt
=D^2(\ell_T-\ln r+3/2)/2$; the remaining terms use
$|\theta_{\rm gf}|\leq2h_{\rm bd}$ and
$|a_{2,\rm gf}|\leq2h_{\rm bd}(\ell_T+3/2)$.
This is an upper bound, not a nonzero error floor or evidence that the
offset attains its bound.

Under the additional pointwise mirror condition
$i_\alpha(u)=f_\alpha(T-u)$, the profiles $P$ and $Q$ are each symmetric
about their own midpoint. Reflection of both integration variables
preserves their ordered superoperator product and changes the sign of
$u-v$, so all odd moments $\mathcal M_p$ vanish. The moment bound then
improves to $|\Delta_n^2z_n|\leq B_{\rm sh}/r_n^2$.
Writing $e_j=\Delta_j^2z_j$ and $S=\sum_{j\geq1}e_j$, the consistently
reanchored coefficients and remainder are
\begin{align}
 b_1^\infty&=b_1+S/2,&
 \delta_{\rm col}^\infty&=\delta_{\rm col}-S/2,\\
 R_{\rm sh}^\infty(n)&=-\sum_{j=1}^{\infty}\min(n,j)e_j,&
 |R_{\rm sh}^\infty(n)|&\leq\frac{B_{\rm sh}}4(2+\ln n),\qquad n\geq1.
\end{align}
The offset bound changes at the same time:
\begin{equation}
 |\delta_{\rm col}^\infty|
 \leq18h_{\rm bd}(\ell_T+3/2)
 +\frac{B_{\rm sh}}2\sum_{j=1}^{\infty}(2j+1)^{-2}.
\end{equation}
The infinite sum is a mathematical reanchoring of the auxiliary log
response; it does not extend the physical finite-band approximation to
an infinite experimental window. The edge error is still evaluated at
each finite node. Approximate mirroring does not inherit the strict
bound without its matching-defect term.

Combining the anchored colored response with the first-white-order term
and reinstating the finite-band and omitted-order corrections gives
\begin{equation}
 E_{\rm gf}(r)=E^*+\delta_{\rm gf}+a_{1,\rm gf}r+a_{2,\rm gf}r^2
       +\theta_{\rm gf}r^2\ln r+R_{\rm gf}(r),
 \qquad r=1,3,5,\ldots .
 \label{eq:sm-fold-four-model}
\end{equation}
Here $\delta_{\rm gf}=\delta_{\rm col}+\Omega$ and
$a_{1,\rm gf}=b_1+a_w$. For $\ell_T\geq\ln3$ and
$\omega_{\rm UV}rT\geq1$ at every node,
\begin{align}
 |\delta_{\rm gf}|&\leq H_{\rm nat}
 =18h_{\rm bd}(\ell_T+3/2)+\frac{\|O\|_\infty}{2}(K_F+K_I),\\
 |R_{\rm gf}(r)|&\leq D_{\rm sh}(r)+\epsilon_{\rm edge}(r)
                          +\epsilon_{\rm high}(r),
 \label{eq:sm-fold-four-certificate}
\end{align}
where $K_X=\int_0^T\|\mathcal L^I_{w,X}(u)\|_{1\to1}\,du$,
$D_{\rm sh}$ is the shape bound above, and $\epsilon_{\rm high}$ is the
mixed-noise omitted-order bound in Eq.~\eqref{sm:high-bound}. Strict mirroring permits the
alternative anchoring and its corresponding bounds, but not removal of
the accompanying constant adjustment. A certified implementation-model
error, when present, is an additional contribution rather than zero by
default.

Equation~\eqref{eq:sm-fold-four-model} is a controlled representation,
not an exact closure of all colored second-order terms in four functions.
Likewise, $a_{2,\rm gf}=-\theta_{\rm gf}(\ell_T+3/2)$ refers to the
declared retained-order auxiliary representation, not the complete mixed
dynamics at arbitrary noise strength. Even in the absence of white
noise, the coefficient $b_1$ need not vanish. Thus pure $1/f$ noise alone
does not justify deleting the linear fold coordinate.

\section{Bias of conventional extrapolation}
\label{sm:conventional-bias}
\subsection{Fixed-node Richardson identity}
Ordinary Richardson extrapolation cancels polynomial response coordinates,
not arbitrary functions of the amplification factor
\cite{Temme2017ErrorMitigation,LiBenjamin2017ActiveErrorMinimization}.
Let $x_j$ denote distinct, fixed physical nodes, with $x=c$ for stretching
or $x=r$ for complete global folding. On the chosen window, write the
retained response as
\begin{equation}
 E(x_j)=E^*+\delta+\sum_{m=1}^{p}a_m x_j^m
       +\theta x_j^2\ln x_j+R(x_j).
 \label{eq:sm-fixed-node-model}
\end{equation}
The protocol-specific coefficients and remainder are those derived above.
For the stretching representation $\delta=0$; for folding, $\delta$ is
the native offset in the declared anchoring convention. Introducing
polynomial columns through degree $p$ specifies the fitted background;
it does not assert that all omitted dynamics are polynomial.

For unbiased node estimates $\widehat E_j$, define
\begin{equation}
 \widehat E_R=\sum_j g_j\widehat E_j,
 \qquad
 \sum_jg_j=1,\qquad
 \sum_jg_jx_j^m=0\quad (m=1,\ldots,p).
 \label{eq:sm-richardson-moments}
\end{equation}
Substitution gives the fixed-node identity
\begin{equation}
 \mathbb E[\widehat E_R]-E^*
 =\delta+\theta M_{2\log}+\sum_jg_jR(x_j),
 \qquad M_{2\log}=\sum_jg_jx_j^2\ln x_j.
 \label{eq:sm-richardson-log-bias}
\end{equation}
For $p\geq2$, the regular quadratic contribution is canceled, including
any cutoff-dependent constant multiplying $x^2$. The ordinary polynomial
conditions do not separately impose $M_{2\log}=0$. The logarithmic
contribution in Eq.~\eqref{eq:sm-richardson-log-bias} is therefore the
specific component left unrepresented by the polynomial model.
For the three-node rule $(1,3,5)$, the logarithmic moment is
$2.729$, quoted as $2.73$ when rounded. This multiplies the isolated
logarithmic coordinate; it is not a replacement for the full finite-band
bias, which also contains the terms shown in
Eq.~\eqref{eq:sm-richardson-log-bias}.

This is a component-wise identity, not an unconditional lower bound on
the total physical bias. In the stretching case, the sufficient
condition
\begin{equation}
 \left|\sum_jg_jR(x_j)\right|
 \leq \epsilon_{\rm rem}|\theta M_{2\log}|,
 \qquad 0\leq\epsilon_{\rm rem}<1,
 \label{eq:sm-richardson-remainder-condition}
\end{equation}
implies
\begin{equation}
 (1-\epsilon_{\rm rem})|\theta M_{2\log}|
 \leq |\mathbb E[\widehat E_R]-E^*|
 \leq(1+\epsilon_{\rm rem})|\theta M_{2\log}|.
 \label{eq:sm-richardson-net-bias}
\end{equation}
The bounds follow from the triangle and reverse-triangle inequalities.
For fixed nodes and a nonzero logarithmic coefficient, they give the
retained-order bias-locking statement when the remainder
condition holds uniformly in the weak-noise limit. Folding also retains
$\delta$ in Eq.~\eqref{eq:sm-richardson-log-bias}; that term cannot be
silently omitted when interpreting a measured intercept.

\subsection{Positive logarithmic moment on a minimal grid}
\label{sm:log-moment}
When the number of nodes is minimal, $M=p+1$, the polynomial constraints
uniquely determine the weights. For distinct positive nodes and $p\geq2$,
the logarithmic moment has the representation
\begin{equation}
 M_{2\log}
 =\int_0^\infty
   \frac{t\prod_{j=0}^{p}x_j}
        {\prod_{j=0}^{p}(t+x_j)}\,dt>0.
 \label{eq:sm-minimal-positive-moment}
\end{equation}
The representation follows by inserting the integral representation of
$\ln x$ into the moment, using the polynomial constraints, and applying
the partial-fraction identity for the minimal interpolation rule. The
integrand is positive, and the integral converges for $p\geq2$.
Consequently, no choice of distinct positive nodes in a minimal ordinary
Richardson rule removes this explicit logarithmic component. With
additional nodes, its cancellation requires an additional constraint and
a full-rank augmented design. This is a response-matched generalized
Richardson construction, rather than a consequence of the polynomial
constraints alone.

The result concerns finite-node rules of the specified type. It neither
provides a node-independent lower bound on total bias nor establishes an
asymptotic claim about every growing node family.
\subsection{Scope of exponential extrapolation}
Repeatable Pauli noise in a gatewise Clifford circuit gives the exact
integer-node product law in Sec.~\ref{sm:clifford-pauli-exp}. The forward
and inverse copy products need not be equal; logical inversion alone
does not justify an ideal-anchored exponential model. We retain the
integer-power identity, without extending a square-root representation
to unspecified signs or zero eigenfactors. This argument does not
apply unchanged to amplitude damping, coherent errors, leakage, or
temporally correlated noise.

Temporally correlated noise falls outside the repeatable-channel
argument. At the retained response order, a quasistatic contribution
produces a quadratic coordinate in the log domain, and the finite-band
$1/f$ response also produces the logarithmic coordinate. A single
exponential, whose log is affine in the amplification factor, does not
represent either additional coordinate. This explains a model mismatch;
it does not establish that exponential extrapolation has the largest
bias for every circuit, memory kernel, and fitting window. Any numerical
ranking must instead refer to a specified circuit, noise model and fitting
window.

\section{Inverse implementations and native constants}
\label{sm:inverse-constants}
The ideal equivalence of folded circuits does not by itself identify the
constant coefficient of a noisy-response fit with $E^*$. This section
states the implementation conditions that remove particular constant
contributions, and separates them from the remainders that remain. Except
where an exact channel identity is stated, we retain white-noise first
order and colored-noise second order. Mixed and higher orders retain the
error accounting of the response expansion.

\subsection{Common-frame matching and repeated gate copies}
\label{sm:common-frame-matching}

Write the target circuit as $U=G_J\cdots G_1$ and let
$P_{j-1}=G_{j-1}\cdots G_1$, with $P_0=I$. The forward and inverse
implementations of gate $j$ have propagators $u_j(t)$ and $v_j(t)$,
equal duration $\tau_j$, and endpoints $G_j$ and $G_j^\dagger$.
Define $\mathcal C_V(X)=VXV^\dagger$. The integrated white-noise
generators in the same local frame are
\begin{align}
 J_{F,j}&=\int_0^{\tau_j}
 \mathcal C_{u_j(t)}^{-1}\mathcal L_{F,j}(t)
 \mathcal C_{u_j(t)}\,dt,\\
 J_{I,j}&=\int_0^{\tau_j}
 \mathcal C_{v_j(t)G_j}^{-1}\mathcal L_{I,j}(t)
 \mathcal C_{v_j(t)G_j}\,dt.
 \label{sm:eq:local-white-integrals}
\end{align}
The factor $G_j$ in the inverse expression is required: an inverse copy
starts from the endpoint of the corresponding forward gate. It cannot
be omitted when comparing physical noise responses.

For one full forward or inverse circuit, the integrated generators are
\begin{equation}
 \mathcal W_X=\sum_j\mathcal C_{P_{j-1}}^{-1}
 J_{X,j}\mathcal C_{P_{j-1}},\qquad X\in\{F,I\}.
 \label{sm:eq:whole-white-integrals}
\end{equation}
Thus $J_{F,j}=J_{I,j}$ for every implemented segment is sufficient for
$\mathcal W_F=\mathcal W_I$, regardless of entanglement in the prefixes
or noncommutation between distinct noise generators. Here a segment
means a segment of the implemented time line. Finite-duration idles
must be included; parallel layers and their correlated couplings cannot
be replaced by a fictitious serial schedule.

The same local comparison controls a larger class of folding schedules.
Start with the forward-copy sequence $(1,+),\ldots,(J,+)$ and repeatedly
replace a contiguous subsequence $V$ by $V\overline V V$, where
$\overline V$ reverses both copy order and copy direction. This includes
global, local, partial and nested folding. The assumptions are that
copies of the same gate in the same direction reuse the same physical
implementation and white-noise generator, and that the colored
environment continues across copy boundaries without being reset.

\paragraph*{Prefix and copy-count identity.}
Every forward copy of gate $j$ starts at ideal prefix $P_{j-1}$, whereas
every inverse copy starts at $G_jP_{j-1}$. If its total copy count is
$n_j$, then its forward and inverse counts are $(n_j+1)/2$ and
$(n_j-1)/2$, respectively. To see this, consider one replacement
$V\mapsto V\overline V V$. The reversed sequence retraces the ideal
prefixes visited by $V$, and the final $V$ repeats the original path.
For each gate, the replacement adds equal numbers of forward and inverse
copies. Induction from the unfurled circuit establishes both the prefix
identity and the forward-minus-inverse count of one.

Let $\rho_j=\mathcal C_{P_{j-1}}(\rho_0)$,
$O_j=\mathcal C_{P_{j-1}}(O_U)$, and define
\begin{align}
 f_j&=\operatorname{Tr}[O_jJ_{F,j}(\rho_j)],&
 i_j&=\operatorname{Tr}[O_jJ_{I,j}(\rho_j)],\\
 \sigma_j&=(f_j+i_j)/2,&\omega_j&=(f_j-i_j)/2.
\end{align}
At white-noise first order, summing the contributions of all copies gives
\begin{equation}
 E_w^\nu=\sum_j n_j^\nu\sigma_j+\Omega,
 \qquad \Omega=\sum_j\omega_j.
 \label{sm:eq:white-copy-count}
\end{equation}
Each forward copy contributes $f_j$ and each inverse copy contributes
$i_j$, because its common-frame prefix is fixed. Multiplication by the
two copy counts proves Eq.~\eqref{sm:eq:white-copy-count}. In global
folding, $n_j^\nu=r$, so the white-noise response is affine in $r$.
Uniform local folding has the same first-order white response, although
its time-correlated contribution need not agree.

The constant $\Omega$ is independent of the folding schedule. Within the
retained-order model allowing independent forward and inverse white
responses, changing schedules alone cannot distinguish $E^*$ from this
unknown additive contribution: the data depend on $E^*+\Omega$.
This is an identifiability statement for that model, not a claim about
arbitrary higher-order identification procedures.

\paragraph*{A sufficient mirror condition.}
Suppose that every gate satisfies
\begin{equation}
 H_{I,j}(t)=-H_{F,j}(\tau_j-t),\qquad
 \mathcal L_{I,j}(t)=\mathcal L_{F,j}(\tau_j-t).
 \label{sm:eq:white-mirror-condition}
\end{equation}
Then $v_j(t)G_j=u_j(\tau_j-t)$: the two sides solve the same
initial-value problem. Substitution in
Eq.~\eqref{sm:eq:local-white-integrals}, followed by reversal of the
integration variable, gives $J_{I,j}=J_{F,j}$. Consequently $\Omega=0$
for every standard folding schedule. This sufficient condition compares
physical controls and white-noise generators, not merely logical gate
endpoints. The use of pulse-based inverse evolution and its dissipative
matching conditions follows the KIK construction of
Ref.~\cite{Henao2023AdaptiveKIK}. Here the copy-count statement specifies
the first-order constant across the declared folding schedules, and the
colored-noise response is treated separately.

For colored couplings, the corresponding local condition is
\begin{equation}
 G_j^\dagger v_j^\dagger(t)B_{I,j,\alpha}(t)v_j(t)G_j
 =u_j^\dagger(\tau_j-t)B_{F,j,\alpha}(\tau_j-t)
 u_j(\tau_j-t).
 \label{sm:eq:colored-local-mirror}
\end{equation}
Conjugating both sides by the common circuit prefix and matching the
forward and inverse time coordinates yields
$b_{I,\alpha}(s)=b_{F,\alpha}(T-s)$ for full blocks. No exchange of the
order of noncommuting coupling operators is used in this step.

Logical self-inversion of CZ is not equivalent to either mirror
condition. For a same-pulse CZ, $v_j=u_j$ and
$J_{I,j}=\mathcal C_{G_j}^{-1}J_{F,j}\mathcal C_{G_j}$.
White-response matching therefore requires
$[J_{F,j},\mathcal C_{G_j}]=0$. Effective diagonal control and fixed
laboratory-frame $Z/ZZ$ dephasing provide one sufficient model. Colored
mirror matching additionally concerns the temporal coupling profile;
diagonality of the ideal endpoint alone does not supply it.

Likewise, an inverse rule established in a two-level control model
cannot be transferred to a multilevel transmon by changing drive signs
alone. Native phase-shifted inverse gates must be characterized in the
physical control model actually used. The single-qubit inverse examples
do not establish an analogous CZ claim.

\subsection{White-noise mismatch bounds and survival observables}
\label{sm:white-mismatch-bounds}

Let $D_j=J_{F,j}-J_{I,j}$ and let
$\Delta O=\lambda_{\max}(O)-\lambda_{\min}(O)$. Trace cyclicity gives
\begin{equation}
 \Omega=\frac12\sum_j\operatorname{Tr}[O_jD_j(\rho_j)].
\end{equation}
Because $D_j(\rho_j)$ is Hermitian and traceless, the spectral midpoint
of $O_j$ can be subtracted without changing this expression. Norm
duality then yields
\begin{equation}
 |\Omega|\leq\frac{\Delta O}{4}
 \sum_j\|D_j(\rho_j)\|_1
 \leq\frac{\Delta O}{4}\sum_j\|D_j\|_{1\to1}.
 \label{sm:eq:white-centered-bound}
\end{equation}
Segments with matched integrated responses contribute zero to this
bound. Their contributions to the $r$-dependent response and higher
orders are not removed.

For approximate control mirrors, a defect certificate is
\begin{align}
 \epsilon_j(t)&\leq\int_0^t
 \|H_{I,j}(s)+H_{F,j}(\tau_j-s)\|_\infty\,ds,\\
 \|D_j\|_{1\to1}&\leq\int_0^{\tau_j}\!dt\,
 \bigl[\|\mathcal L_{I,j}(t)-\mathcal L_{F,j}(\tau_j-t)\|_{1\to1}
 +4\|\mathcal L_{F,j}(\tau_j-t)\|_{1\to1}\epsilon_j(t)\bigr].
 \label{sm:eq:implementation-defect}
\end{align}
Here $\epsilon_j(t)$ bounds the operator-norm distance between
$v_j(t)G_j$ and $u_j(\tau_j-t)$. The propagator difference follows from
the Duhamel identity; both flanking propagators are unitary, so no
Gronwall factor is needed. The second bound separates the generator
mismatch and bounds the two conjugation differences. It is a
model-dependent certificate, not a calibration that is inferred from
logical gate labels.

\paragraph*{Projector survival.}
For pure $\rho_0$ and $O_U=\rho_0$, the Lindblad dissipator obeys
\begin{equation}
 \operatorname{Tr}\{\rho_j\mathcal D[L](\rho_j)\}
 =|\langle L\rangle_{\rho_j}|^2
 -\langle L^\dagger L\rangle_{\rho_j},\qquad
 \operatorname{Tr}\{\rho_j[H,\rho_j]\}=0,
\end{equation}
where $\mathcal D[L](\rho)=L\rho L^\dagger-\{L^\dagger L,\rho\}/2$.
Thus the first-order offset is half the difference between the forward
and inverse integrated jump variances. The coherent contribution
vanishes in this observable, but the dissipative difference need not.

For a same-pulse effective diagonal CZ, additionally assume identical
local white generators for the two directions. Consider one relaxation
jump on qubit $q$, with partner $k$. Since
$G\sigma_-^{(q)}G=\sigma_-^{(q)}Z_k$, the common-frame forward jump
can be written $A(t)+B(t)$ and the inverse jump $A(t)-B(t)$, with
$A$ and $B$ in the two partner-$Z_k$ sectors. Their $L^\dagger L$
terms agree. The resulting offset density is
\begin{equation}
 d\Omega_{j,q}=2\gamma_q(t)
 \operatorname{Re}\{\langle A(t)\rangle\langle B(t)\rangle^*\}\,dt.
\end{equation}
If $p_{ab}$ are the pair's computational-basis probabilities after
tracing over the rest of the register, Cauchy--Schwarz gives
$|\langle A\rangle|\leq\sqrt{p_{00}p_{10}}$ and
$|\langle B\rangle|\leq\sqrt{p_{01}p_{11}}$.
Their product is at most $1/16$, by normalization and the
arithmetic--geometric mean inequality. Hence
\begin{equation}
 |\Omega_{j,q}|\leq\frac18\int_0^{\tau_j}\gamma_q(t)\,dt.
 \label{sm:eq:cz-projector-bound}
\end{equation}

For the contrast observable $2P-1$, the normalization changes this bound to
\begin{equation}
 |\Omega_{j,q}^{\rm contrast}|\leq
 \frac14\int_0^{\tau_j}\gamma_q(t)\,dt.
 \label{sm:eq:cz-contrast-bound}
\end{equation}
Its normalization follows from $O=2\Pi-I$: the identity part contributes
zero under the trace-preserving first-order correction, while the
projector part is multiplied by two.

These bounds apply only to the stated relaxation channel and effective
CZ model. They do not cover an arbitrary multiqubit Pauli observable,
leakage, crosstalk, or a mismatched forward/inverse generator. For the
Pauli observables in the application, only the applicable general bound
in Eq.~\eqref{sm:eq:white-centered-bound} can be invoked. A gate-time
rate $\gamma_q(t)$ cannot be replaced by an idle $T_1^{-1}$ without
supporting characterization.

\subsection{The infrared-enhanced constant}
\label{sm:infrared-constant}

The quasi-static response separates integrated coupling mismatch from
time ordering. Define
\begin{align}
 Q_{X,\alpha}&=\int_0^T b_{X,\alpha}(t)\,dt,&
 Q_{\pm,\alpha}&=(Q_{F,\alpha}\pm Q_{I,\alpha})/2,\\
 \mathcal B(Q,R)&=\sum_{\alpha\beta}\mathsf A_{\alpha\beta}
 \operatorname{ad}_{Q_\alpha}\operatorname{ad}_{R_\beta},\\
 \mathcal C_X&=\frac12\sum_{\alpha\beta}\mathsf A_{\alpha\beta}
 \int_0^Tdt\int_0^t ds\,
 [\operatorname{ad}_{b_{X,\alpha}(t)},
  \operatorname{ad}_{b_{X,\beta}(s)}],
 \label{sm:eq:ordered-static-response}
\end{align}
where $\operatorname{ad}_B(X)=[B,X]$. The ordered triangular integral
within block $X$ equals
$\mathcal B(Q_X,Q_X)/2+\mathcal C_X$: the full time square is the
sum of the two triangular orderings, and their difference is the
commutator integral in Eq.~\eqref{sm:eq:ordered-static-response}.

Counting the cross-block pairs gives the exact static identity
\begin{equation}
 \mathcal K_r^{\rm stat}=\frac12
 \mathcal B(rQ_++Q_-,rQ_++Q_-)
 +\frac r2(\mathcal C_F+\mathcal C_I)
 +\frac12(\mathcal C_F-\mathcal C_I).
 \label{sm:eq:static-response-identity}
\end{equation}
The two mixed block orientations occur equally often, so their
antisymmetric contributions cancel. Taking the observable contraction
defines $s_2r^2+s_1r+s_0$, with
\begin{align}
 s_2&=\frac12\operatorname{Tr}
 [O_U\mathcal B(Q_+,Q_+)\rho_0],\\
 s_1&=\frac12\operatorname{Tr}
 [O_U\{\mathcal B(Q_+,Q_-)+\mathcal B(Q_-,Q_+)
       +\mathcal C_F+\mathcal C_I\}\rho_0],\\
 s_0&=\frac12\operatorname{Tr}
 [O_U\{\mathcal B(Q_-,Q_-)+\mathcal C_F-\mathcal C_I\}\rho_0].
 \label{sm:eq:static-coefficients}
\end{align}
Strict colored mirror matching gives $Q_-=0$ and
$\mathcal C_I=-\mathcal C_F$, but it does not in general set
$\mathcal C_F$ to zero.

For the logarithmic covariance kernel, define the cutoff-independent
shape integral
\begin{equation}
 L_r=\sum_{\alpha\beta}\mathsf A_{\alpha\beta}
 \int_0^{rT}\!dt\int_0^t\!ds\,
 \left[-\ln\frac{t-s}{T}\right]
 \operatorname{Tr}[O_U\operatorname{ad}_{b_{r,\alpha}(t)}
 \operatorname{ad}_{b_{r,\beta}(s)}\rho_0].
 \label{sm:eq:shape-integral}
\end{equation}
The diagonal is interpreted as an integrable improper limit. The
colored correction in the auxiliary logarithmic-kernel model is
\begin{equation}
 \Delta E_{\log}^{(2)}(r)=-\frac{\lambda^2}{\pi}
 [\ell_T(s_2r^2+s_1r+s_0)+L_r],
 \qquad \theta_{\rm gf}=\frac{\lambda^2s_2}{\pi}.
\end{equation}
Under the fixed native anchoring used in the four-function response,
substitution gives
\begin{align}
 \delta_{\rm col}&=-\frac{\lambda^2\ell_T}{\pi}s_0
                  +\delta_{\rm shape},\\
 \delta_{\rm shape}&=\frac{\lambda^2}{2\pi}(L_3-3L_1)
              +\frac92\theta_{\rm gf}(\ln3-1).
 \label{sm:eq:infrared-offset-split}
\end{align}
The infrared logarithm therefore multiplies $s_0$, not the whole
within-block response. This decomposition is an identity for the
declared anchoring; it is not an additional fitting assumption.

To state the cancellation condition, write
$\mathcal C_X=\operatorname{ad}_{J_X^{\rm ord}}$, where
\begin{equation}
 J_X^{\rm ord}=\frac12\sum_{\alpha\beta}\mathsf A_{\alpha\beta}
 \int_0^Tdt\int_0^t ds\,
 [b_{X,\alpha}(t),b_{X,\beta}(s)].
\end{equation}
This follows from the Jacobi identity. With
$q_\alpha=\|Q_{-,\alpha}\|_\infty$, trace cyclicity and norm duality
give
\begin{equation}
 |s_0|\leq2\|O\|_\infty
 \sum_{\alpha\beta}|\mathsf A_{\alpha\beta}|q_\alpha q_\beta
 +\frac12\|[\rho_0,O_U]\|_1
 \|J_F^{\rm ord}-J_I^{\rm ord}\|_\infty.
 \label{sm:eq:s0-structural-bound}
\end{equation}
Integrated matching $Q_F=Q_I$ removes the first term. The commutation
condition $[\rho_0,O_U]=0$ removes the second. Together they imply
$s_0=0$ and hence remove the infrared-enhanced constant in
Eq.~\eqref{sm:eq:infrared-offset-split}. Pure-state projector survival
satisfies the commutation condition, but survival alone does not imply
integrated response matching.

The result does not remove $\delta_{\rm shape}$, finite-band corrections
or omitted noise orders. With the normalization in
Eq.~\eqref{eq:sm-fold-bound-scale}, its shape bound is
\begin{equation}
 |\delta_{\rm shape}|\leq(18\ln3-7)h_{\rm bd},
 \qquad
 |\delta_{\rm col}|\leq
 \frac{\lambda^2|\ell_T|}{\pi}|s_0|+(18\ln3-7)h_{\rm bd},
 \label{sm:eq:colored-native-bound}
\end{equation}
where $\|b_{X,\alpha}(t)\|_\infty\leq\beta_\alpha$.
This follows by bounding the shape integral with
$\int_0^r(r-x)|\ln x|\,dx$ and applying the fixed-anchor identity.
It is a certificate, not a proof that the actual residual is small
compared with the logarithmic contribution.

\paragraph*{Comparing the two logarithmic coefficients.}
Take the same forward implementation as the reference for ideal stretching.
Its ordered static integral gives
\begin{equation}
 \theta_{\rm ps}=\frac{\lambda^2}{\pi}
 \Tr\!\left[O_U\left\{\frac12\mathcal B(Q_F,Q_F)
 +\mathcal C_F\right\}\rho_0\right].
\end{equation}
If $Q_I=Q_F$, the folding coefficient in
Eq.~\eqref{eq:sm-fold-theta} therefore satisfies
\begin{equation}
 \theta_{\rm ps}-\theta_{\rm gf}
 =\frac{\lambda^2}{\pi}\Tr(O_U\mathcal C_F\rho_0).
 \label{sm:protocol-coefficient-comparison}
\end{equation}
Since $\mathcal C_F$ is a commutator superoperator, trace cyclicity sets
the right-hand side to zero when $[O_U,\rho_0]=0$. Integrated matching and
this commutation condition are sufficient for a shared coefficient;
neither survival alone nor logical inversion alone establishes the equality.
These additional conditions compare the two protocols and are not premises
of the general fold counting.

\subsection{Higher orders and the Clifford--Pauli special case}
\label{sm:higher-orders-pauli}

Mirror matching does not remove every constant term at higher order.
For a repeated block generator linear in a strength parameter
$\varepsilon$, let the common-frame forward and inverse block maps be
$\mathcal A(\varepsilon)$ and $\mathcal B(\varepsilon)$. Suppose
$\widetilde{\mathcal L}_{n,I}(s)=
\widetilde{\mathcal L}_{n,F}(T-s)$.
For global folding,
$\Phi_r=\mathcal A(\mathcal B\mathcal A)^n$ with $r=2n+1$.
Writing $E_r=\sum_pE_p(r)\varepsilon^p$, the parity relation states that
$E_p(r)$ is a polynomial of degree at most $p$ and
\begin{equation}
 E_p(-r)=(-1)^pE_p(r).
 \label{sm:eq:response-parity}
\end{equation}
Thus odd orders have no constant term, whereas even orders can contain
one, including white-noise second order.

\textit{Proof.}
Reversing a time-ordered exponential and changing the generator sign
gives $\mathcal B(\varepsilon)=\mathcal A(-\varepsilon)^{-1}$.
Since $\mathcal B\mathcal A=I+O(\varepsilon)$, expansion of
$\exp[n\log(\mathcal B\mathcal A)]$ gives a coefficient of order
$\varepsilon^p$ that is a polynomial in $n$ of degree at most $p$.
The continuation $n\mapsto-n-1$, or $r\mapsto-r$, then gives
$\Phi_{-r}(\varepsilon)=\Phi_r(-\varepsilon)$.
Comparison of coefficients proves Eq.~\eqref{sm:eq:response-parity}.
This argument requires repeated block generators; it cannot be applied
directly to temporally correlated noise realizations that differ from
block to block.

\paragraph*{An exact product law for Pauli channels.}
\label{sm:clifford-pauli-exp}
Consider a circuit whose individual ideal gates are Clifford and whose
observable is a signed Pauli string. Suppose all noise consists of
fixed-position Pauli channels before and after each gate, identical for
all copies of a given gate and direction, with no drift. This model
excludes amplitude damping, coherent errors, leakage and temporally
correlated noise. If
$\mathcal N(\rho)=\sum_Qp_QQ\rho Q$, define its Pauli eigenfactor by
\begin{equation}
 f(\mathcal N,S)=\sum_Qp_Q(-1)^{[Q,S]\ne0},
 \qquad \mathcal N^\dagger(S)=f(\mathcal N,S)S.
\end{equation}
Let $S_j=P_jO_UP_j^\dagger$, and set
\begin{align}
 \Pi_F&=\prod_j
 f(\mathcal N_j^{F,\mathrm{pre}},S_{j-1})
 f(\mathcal N_j^{F,\mathrm{post}},S_j),\\
 \Pi_I&=\prod_j
 f(\mathcal N_j^{I,\mathrm{pre}},S_j)
 f(\mathcal N_j^{I,\mathrm{post}},S_{j-1}).
\end{align}
The integer-node response is exactly
\begin{equation}
 E(2n+1)=E^*\Pi_F^{n+1}\Pi_I^n.
 \label{sm:eq:clifford-product-law}
\end{equation}

\textit{Proof.}
Back-propagate the measured Pauli string in the Heisenberg picture.
Clifford conjugation preserves a signed Pauli string, and each Pauli
channel multiplies it by its eigenfactor. The prefix identity fixes the
Pauli strings seen by every forward and inverse copy. There are $n+1$
forward copies and $n$ inverse copies of each gate, giving
Eq.~\eqref{sm:eq:clifford-product-law}.

Equation~\eqref{sm:eq:clifford-product-law} is stated in integer powers
and does not require a choice of square-root branch. We do not use an
unrestricted real-exponential continuation of that identity.
Forward/inverse product matching, $\Pi_F=\Pi_I$, is an additional
physical condition. A sufficient channel-placement condition is
$\mathcal N_j^{I,\mathrm{pre}}=\mathcal N_j^{F,\mathrm{post}}$ and
$\mathcal N_j^{I,\mathrm{post}}=\mathcal N_j^{F,\mathrm{pre}}$.
Identical post-gate channels alone need not satisfy it, because the two
directions generally expose different Pauli strings to the channel.

This special case motivates a log-domain treatment of the Markovian
contribution under its stated assumptions. It is not an exact theorem
for the mixed-noise setting: temporal correlations are excluded
from the product law and must be treated through the response kernel.
Nor does this channel identity establish the finite-sample properties
or the complete error certificate of PE\_log.

\subsection{Model-assisted closure and identifiability}
\label{sm:model-assisted-closure}

This extension uses additional model information and is separate from
the main four-function PE\_log protocol. Assume that the forward and
inverse controls and coupling operators are known for every implemented
gate. Write the white generator as
$\mathcal L_w=\sum_c g_c\overline{\mathcal L}_c$ and the colored
spectral matrix as
$\lambda^2\mathsf A=\sum_\mu p_\mu\overline{\mathsf A}^{(\mu)}$,
with known generator and spectral shapes but unknown amplitudes
$g_c,p_\mu\geq0$. Each colored mode has unknown finite cutoffs and
$\ell_\mu=-\gamma_{\mathrm E}-\ln(\omega_{\mathrm{IR},\mu}T)$.
The shapes must be shared between the compared schedules as specified
by the model; logical gate names alone do not provide them.

For schedule $\nu$, denote its physical duration by $D_\nu$ and ideal
cumulative propagator by $V_\nu(t)$. Define the model-computable
response columns
\begin{align}
 W_c^\nu={}&\int_0^{D_\nu}\!dt\,
 \operatorname{Tr}[O_U\mathcal C_{V_\nu(t)}^{-1}
 \overline{\mathcal L}_c(t)\mathcal C_{V_\nu(t)}\rho_0],\\
 S_\mu^\nu={}&-\frac1\pi\sum_{\alpha\beta}
 \overline{\mathsf A}_{\alpha\beta}^{(\mu)}
 \int_0^{D_\nu}\!dt\int_0^t\!ds\,
 \operatorname{Tr}[O_U\operatorname{ad}_{b_{\nu,\alpha}(t)}
 \operatorname{ad}_{b_{\nu,\beta}(s)}\rho_0],\\
 Y_\mu^\nu={}&-\frac1\pi\sum_{\alpha\beta}
 \overline{\mathsf A}_{\alpha\beta}^{(\mu)}
 \int_0^{D_\nu}\!dt\int_0^t\!ds\,
 \left[-\ln\frac{t-s}{T}\right]
 \operatorname{Tr}[O_U\operatorname{ad}_{b_{\nu,\alpha}(t)}
 \operatorname{ad}_{b_{\nu,\beta}(s)}\rho_0].
 \label{sm:eq:exact-response-columns}
\end{align}
These are noise-response integrals, not measured node values or
estimates obtained by substituting an unknown ideal value. Inserting
the logarithmic covariance representation in the second-order response
gives
\begin{equation}
 E_\nu=E^*+\sum_c g_cW_c^\nu
 +\sum_\mu(q_\mu S_\mu^\nu+p_\mu Y_\mu^\nu)+R_\nu,
 \qquad q_\mu=p_\mu\ell_\mu.
 \label{sm:eq:known-shape-model}
\end{equation}
The exact columns retain the within-order control shape; $R_\nu$
contains finite-band replacement and higher-order errors. In contrast,
replacing these columns by the four-function reduction can leave a
shape residual at colored-noise second order.

Collect the noise columns into $X_n$, evaluated on $N_s$ selected
schedules, and let $\Pi_n$ be the orthogonal projector onto
$\operatorname{col}X_n$. Define
\begin{equation}
 \kappa_{\rm id}=\frac{\|(I-\Pi_n)\mathbf1\|_2}{\sqrt{N_s}}.
 \label{sm:eq:identifiability-margin}
\end{equation}
If $\kappa_{\rm id}>0$, a linear estimator can preserve the ideal
column and annihilate every declared noise column. Its minimum
Euclidean-norm weights are
\begin{equation}
 \mathbf z=(I-\Pi_n)\mathbf1,\qquad
 \mathbf w=\frac{\mathbf z}{\|\mathbf z\|_2^2},\qquad
 \|\mathbf w\|_2=\frac1{\kappa_{\rm id}\sqrt{N_s}},
 \label{sm:eq:closure-weights}
\end{equation}
and obey
$\mathbf1^{\mathsf T}\mathbf w=1$,
$X_n^{\mathsf T}\mathbf w=0$.
For unbiased node estimates and certificates $|R_\nu|\leq B_\nu$,
\begin{equation}
 \mathbb E\!\left[\sum_\nu w_\nu\widehat E_\nu\right]-E^*
 =\sum_\nu w_\nu R_\nu,
 \qquad
 \left|\mathbb E\!\left[\sum_\nu w_\nu\widehat E_\nu\right]-E^*\right|
 \leq\sum_\nu|w_\nu|B_\nu.
 \label{sm:eq:exact-closure-certificate}
\end{equation}

\textit{Proof.}
$\mathbf z$ is orthogonal to every noise column, and
$\mathbf1^{\mathsf T}\mathbf z=\|\mathbf z\|_2^2>0$.
Consequently Eq.~\eqref{sm:eq:closure-weights} satisfies all moment
conditions. The difference between this solution and another feasible
solution is orthogonal to both $\mathbf1$ and the noise-column space;
the Pythagorean identity establishes the minimum norm. Substitution of
Eq.~\eqref{sm:eq:known-shape-model} proves
Eq.~\eqref{sm:eq:exact-closure-certificate}. If
$\kappa_{\rm id}=0$, the constant column lies in the noise-column
space and these linear moment conditions have no solution.

This construction needs neither the unknown amplitudes nor the
cutoffs to determine its weights. A numerical residual certificate
nevertheless needs justified ranges for the actual amplitudes and
cutoffs. Control-model errors produce additional column mismatch;
they are not removed by labelling the correction model-assisted.
For large circuits, computing the required response columns can itself
be costly. These are conditions of the extension, not assumed
requirements of the four-function estimator itself.

For global folding, $W_c(r)=r\bar\sigma_c+\Omega_c$.
Two independent pairs $(\bar\sigma_c,\Omega_c)$ can span a constant
column, giving $\kappa_{\rm id}=0$ on every global node set. When all
$\Omega_c=0$, this particular white-noise obstruction is absent, but
the full criterion still includes the colored columns. A change of
folding schedule can improve the rank of a specified known-shape model;
it does not by itself remove an arbitrary unknown $\Omega$.

\subsection{Finite-band constants are not canceled by mirror matching}
\label{sm:finite-band-constant}

One remaining constant is visible in the ultraviolet edge correction.
Let $K(t,s)$ denote the scalar double-commutator kernel, including its
spatial contraction, and define
$k_\nu(\tau)=\int_\tau^{D_\nu}K(t,t-\tau)\,dt$.
Suppose $k_\nu\in C^1$, its first derivative is absolutely continuous,
and $|k_\nu''|\leq M_2$. For $\omega_{\rm UV}D_\nu\geq4$ the
ultraviolet part of the finite-band correction is
$-(\lambda^2/\pi)\int_0^{D_\nu}
\operatorname{Ci}(\omega_{\rm UV}\tau)k_\nu(\tau)d\tau$.
Two integrations by parts give
\begin{align}
 &\left|\int_0^{D_\nu}\operatorname{Ci}(\omega_{\rm UV}\tau)
 k_\nu(\tau)d\tau-\frac{k_\nu'(0^+)}{2\omega_{\rm UV}^2}\right|
 \notag\\
 &\qquad\leq\frac1{\omega_{\rm UV}^2}
 \left[\frac{2|k_\nu'(D_\nu)|}{\omega_{\rm UV}D_\nu}
 +\frac{2M_2}{\omega_{\rm UV}}
 \left(1+\ln\frac{\omega_{\rm UV}D_\nu}{4}\right)\right].
 \label{sm:uv-tail-bound}
\end{align}
For completeness, define the twice-integrated cosine integral by
$\mathcal E(u)=u^2\int_u^\infty v^{-3}\cos v\,dv$.
The bounds $|\mathcal E(u)|\leq\min(1/2,2/u)$ and $k_\nu(D_\nu)=0$
give the exact identity
\begin{align}
 \int_0^{D_\nu}\operatorname{Ci}(b\tau)k_\nu(\tau)d\tau
 =\frac1{b^2}\left[\frac{k_\nu'(0^+)}2
 -\mathcal E(bD_\nu)k_\nu'(D_\nu)
 +\int_0^{D_\nu}\mathcal E(b\tau)k_\nu''(\tau)d\tau\right],
 \qquad b=\omega_{\rm UV}.
\end{align}
Splitting the last integral where $b\tau=4$ proves
Eq.~\eqref{sm:uv-tail-bound}. Thus the leading ultraviolet contribution is
\begin{equation}
 -\frac{\lambda^2}{2\pi\omega_{\rm UV}^2}k_\nu'(0^+),
 \qquad
 k_\nu'(0^+)=-K(0,0)
 -\int_0^{D_\nu}\!dt\,\partial_sK(t,s)\big|_{s=t},
 \label{sm:eq:uv-constant}
\end{equation}
with the finite-band remainder controlled separately. For gatewise
control and coupling mirrors, inverse copies contribute the negatives
of the forward-copy contributions to the derivative integral. The
forward-minus-inverse count is one for every gate, and the initial
kernel value is common to all schedules. Hence
$k_\nu'(0^+)=k_1'(0^+)$: the leading ultraviolet term is a common
constant, retained by any weights summing to one.

This observation completes the distinction between the cancellation
claims. Matching removes the white-noise first-order constant, while
integrated colored matching and $[\rho_0,O_U]=0$ remove the
infrared-enhanced $s_0$ contribution. Neither statement removes the
shape residual of the reduced basis, this finite-band constant,
allowed even-order constants, or the full higher-order remainder.

\section{Response-matched estimators}
\label{sm:estimators}

\subsection{Moment constraints and the linear certificate}
For fixed positive nodes, define
\begin{equation}
 X_{k\cdot}=(1,x_k,x_k^2,x_k^2\ln x_k),\qquad
 \boldsymbol w^{\mathsf T}=\boldsymbol e_1^{\mathsf T}X^+,
 \label{eq:sm-design-and-weights}
\end{equation}
where $X^+$ is the unweighted Moore--Penrose pseudoinverse and
$\boldsymbol e_1$ selects the intercept. Complete global folding
uses distinct positive odd integers $x_k=r_k$; pulse stretching uses
the physical factors $x_k=c_k$.

The design has full column rank on at least four distinct positive nodes.
If a function $p_2(x)+b x^2\ln x$ had four distinct positive zeros,
Rolle's theorem applied three times would require a zero of $2b/x$.
Thus $b=0$, and a quadratic with four zeros is identically zero.
Additional distinct nodes preserve column independence. Consequently
\begin{equation}
 \sum_kw_k=1,\qquad
 \sum_kw_kx_k=\sum_kw_kx_k^2=\sum_kw_kx_k^2\ln x_k=0.
 \label{eq:sm-response-moments}
\end{equation}
The nodes need not be $(1,3,5,7)$. The moment constraints determine
weights without knowing $E^*$, the noise amplitudes, or the response
coefficients. They do not determine whether the physical remainder
is negligible on a particular device.

For unbiased node estimates, the linear-domain estimator
$\widehat E_{\mathrm{PE,lin}}=\sum_kw_k\widehat E(x_k)$ obeys
\begin{equation}
 \mathbb E[\widehat E_{\mathrm{PE,lin}}]-E^*
 =\delta+\sum_kw_kR(x_k).
 \label{eq:sm-linear-pe-identity}
\end{equation}
For folding, $|\delta|\leq H_{\rm nat}$ and $|R(r_k)|\leq D(r_k)$ give
\begin{equation}
 \left|\mathbb E[\widehat E_{\mathrm{PE,lin}}]-E^*\right|
 \leq H_{\rm nat}+\|\boldsymbol w\|_1D_{\max},
 \qquad D_{\max}=\max_kD(r_k).
 \label{eq:sm-linear-pe-certificate}
\end{equation}
No independence between node estimates is needed for this expectation
identity. The bound retains native offsets, control-shape corrections,
finite-band effects, and omitted noise orders. Removal of the logarithmic
coordinate is not elimination of every term at colored-noise second order.

\subsection{Log-domain response and PE\_log}
Let $E^*\ne0$ and assume that the population responses lie on its fixed
nonzero sign branch. The local Taylor expansion gives
\begin{equation}
 \ln|E|=\ln|E^*|+\frac{E-E^*}{E^*}
 -\frac{(E-E^*)^2}{2(E^*)^2}+\cdots.
 \label{eq:sm-log-retained-expansion}
\end{equation}
Through first white-noise order and second colored-noise order, the
response coordinates are therefore the same as in the linear domain,
with coefficients divided by $E^*$. Squaring the correction gives white
second order, mixed orders, and colored fourth order, outside the
declared truncation. Shape and finite-band contributions remain.

For positive measured inputs $y_k$, define
\begin{equation}
 \widehat E_{\mathrm{PE,log}}
 =\exp\!\left(\sum_kw_k\ln y_k\right).
 \label{eq:sm-log-pe-estimator}
\end{equation}
If a negative sign branch is independently specified, the analogous
definition uses $\ln|y_k|$ and restores that fixed sign after exponentiation.
A zero ideal expectation has no such logarithmic expansion; the
linear-domain estimator does not require this nonzero-sign assumption.
A nonpositive input invalidates the positive-branch definition. It must
not be silently deleted or replaced by an absolute value in that protocol.

An exponential attenuation becomes linear in the log domain. The exact
integer-node Clifford--Pauli product in Sec.~\ref{sm:clifford-pauli-exp}
explains a special case. Under positive nonzero channel-copy products,
its logarithm has the algebraic form
\[
 \ln|E(2n+1)|=\ln|E^*|+(n+1)\ln\Pi_F+n\ln\Pi_I .
\]
Forward--inverse matching is required to identify the intercept with
$\ln|E^*|$. This does not extend the special channel theorem to
amplitude damping, coherent errors, leakage, or correlated classical
fields.

PE\_log cancels the explicitly derived logarithmic coordinate at retained
response order. It is not equal to PE\_lin at finite noise or finite sample
size. Equation~\eqref{eq:sm-linear-pe-certificate} is a certificate for
the linear estimator only; taking a logarithm and exponentiating introduces
additional deterministic and sampling effects. Near-zero signals also
make the transformation ill-conditioned. Neither exact finite-sample
unbiasedness nor a general statistical advantage is asserted.

\subsection{Explicit weights and a conditional bias comparison}
For the four-node family $r_k=Ls_k$, with
$\boldsymbol s=(1,3,5,7)^{\mathsf T}$ and $L$ positive and odd,
the weights are independent of $L$, because
\[
 r_k^2\ln r_k=L^2s_k^2\ln s_k+L^2(\ln L)s_k^2.
\]
All four nodes are physical complete-fold nodes. For $L>1$ the
sampling grid need not contain the original circuit $r=1$.

The quadratic ordinary-least-squares intercept rule on the same four
nodes has weights
\[
 \boldsymbol q=\frac1{80}(123,-19,-51,27)^{\mathsf T},\qquad
 \boldsymbol v=(-1,3,-3,1)^{\mathsf T}.
\]
The vector $\boldsymbol v$ spans the null space of the first three
moment conditions, and $\boldsymbol q^{\mathsf T}\boldsymbol v=0$.
Thus $\boldsymbol q$ is the minimum-norm solution of those conditions.
Define
\begin{align}
 \beta&=\boldsymbol q^{\mathsf T}(\boldsymbol s^2\ln\boldsymbol s)
 =\frac{-171\ln3-1275\ln5+1323\ln7}{80}>0,\\
 h&=\boldsymbol v^{\mathsf T}(\boldsymbol s^2\ln\boldsymbol s)
 =27\ln3-75\ln5+49\ln7>0 .
\end{align}
Here $h$ is a third difference of $f(s)=s^2\ln s$, and
\[
 h=\int_0^2\!\int_0^2\!\int_0^2
 f'''(1+x+y+z)\,dx\,dy\,dz>0.
\]
The stated sign of $\beta$ can be checked with rigorous logarithm
bounds from the convergent series
\[
 \ln x=2\sum_{m=0}^{M}\frac{z^{2m+1}}{2m+1}+\mathcal E_M,\quad
 z=\frac{x-1}{x+1},\quad
 |\mathcal E_M|\leq
 \frac{2|z|^{2M+3}}{(2M+3)(1-z^2)}.
\]
The four-function weights are
\begin{equation}
 \boldsymbol w=\boldsymbol q-\frac{\beta}{h}\boldsymbol v .
 \label{eq:sm-explicit-weights}
\end{equation}
These are algebraic interpolation weights, not fitted simulation outputs.

Writing $b$ for expectation-value bias, the two linear estimators satisfy
\begin{align}
 b_{\mathrm{PE,lin}}&=\delta+\sum_kw_kR(Ls_k),\\
 b_Q&=\delta+\theta\beta L^2+\sum_kq_kR(Ls_k).
 \label{eq:sm-linear-comparison-biases}
\end{align}
With $D_L=\max_kD(Ls_k)$, a sufficient condition for strict bias reduction is
\begin{equation}
 |\theta|\beta L^2>
 2H_{\rm nat}+(\|\boldsymbol w\|_1+\|\boldsymbol q\|_1)D_L .
 \label{eq:sm-linear-bias-condition}
\end{equation}
Indeed, the reverse-triangle inequality gives
$|b_Q|\geq|\theta|\beta L^2-H_{\rm nat}-\|\boldsymbol q\|_1D_L$,
whereas $|b_{\mathrm{PE,lin}}|\leq
H_{\rm nat}+\|\boldsymbol w\|_1D_L$. The displayed condition makes
the former bound strictly larger, without assuming favorable signs
for the offset or residuals. This is a conditional comparison of
deterministic biases, not a PE\_log theorem or a finite-sample
mean-squared-error guarantee.

\subsection{A nonempty bias-improvement region}
The sufficient condition above has a nonempty domain within the stated
physical model. Fix the complete forward and inverse controls, the input,
observable, coupling operators, and spectral-matrix shape, with
$|\theta|/h_{\rm bd}=c_\theta>0$. Consider a family of distinct physical
designs indexed by positive odd $L\to\infty$:
\begin{equation}
 \omega_{\rm IR}T=L^{-2},\qquad
 \omega_{\rm UV}T=B_0\geq1,\qquad
 \lambda^2\mathcal B_AT^2=c_\lambda L^{-4},\qquad
 \gamma T=c_wL^{-4}.
 \label{eq:sm-nonempty-family}
\end{equation}
Here $B_0,c_\lambda,c_w$ are fixed positive constants and $\gamma$
scales a fixed white-generator profile. Within each design, all four
nodes share the same spectrum and both noise strengths. The parameters
vary between designs, not between nodes of one extrapolation.

Now $h_{\rm bd}=O(L^{-4})$,
$\ell_T=2\ln L-\gamma_{\rm E}$, and $r\leq7L$.
The term canceled by PE\_lin but retained by the quadratic rule is
$|\theta|\beta L^2=\Theta(L^{-2})$.
The colored and white native-constant bounds are
$O(L^{-4}\ln L)$ and $O(L^{-4})$, respectively.
The generic, nonmirror shape bound is $O(L^{-3}\ln L)$.
The infrared and ultraviolet parts of the edge bound become
\[
 \frac{h_{\rm bd}}{12}(\omega_{\rm IR}T)^2r^4=O(L^{-4}),\qquad
 \frac{8h_{\rm bd}r}{B_0}[1+\ln(B_0r)]
 =O(L^{-3}\ln L/B_0).
\]
The total white norm is $K_w=O(L^{-3})$, while the finite-band Gaussian
variance gives $Z=O(L^{-2}\ln L)$. Hence
$e^{K_w+Z}-1-K_w-Z=O(L^{-4}\ln^2L)$.
Every residual bound is asymptotically smaller than the
$\Theta(L^{-2})$ logarithmic bias component. The weights and $\beta$
are fixed, so Eq.~\eqref{eq:sm-linear-bias-condition} holds at some
finite odd $L$ and for all sufficiently large members of this family.
Both noise components remain nonzero; mirror matching is not used.

This is an existence result, not a large-depth prescription for a fixed
device. Holding its cutoffs and strengths fixed while increasing $L$
eventually invalidates the small-remainder regime. The construction
neither selects experimental nodes nor establishes a finite-budget
advantage.

\subsection{An entangling-circuit witness}
A finite-duration two-qubit circuit shows that $c_\theta>0$ is attainable
without restricting the structural theorem to a special circuit:
\[
 \rho_0=|00\rangle\langle00|,\qquad
 U=\mathrm{CZ}(H_{\rm Had}\otimes H_{\rm Had}),\qquad
 O=X_1Z_2 .
\]
Let the simultaneous Hadamard layer and CZ take durations $T_H>0$
and $T_C>0$, with $T=T_H+T_C$. All gates and inverse gates participate
in complete folding. Take laboratory coupling $B=Z_1/2$ and scalar
spectral weight $A>0$. For this witness only, use the effective
diagonal CZ Hamiltonian
$H_{\rm CZ}(t)=\chi(t)|11\rangle\langle11|$ with
$\int_0^{T_C}\chi(t)\,dt=\pi$; the same CZ pulse is its logical inverse.

Since $O_U=Z_1$, the ideal expectation is one. In the common frame,
both CZ intervals have coupling $X_1/2$. Each finite Hadamard interval
has coupling norm $1/2$, so
\[
 Q_F+Q_I=T_CX_1+\Delta Q,\qquad \|\Delta Q\|_\infty\leq T_H .
\]
The leading double-commutator contraction is $4T_C^2$.
The cross terms are bounded in total by $8T_CT_H$, and the remaining
term by $4T_H^2$. Inserting $Q_+=(Q_F+Q_I)/2$ gives
\begin{equation}
 \theta\geq\frac{\lambda^2A}{2\pi}
 (T_C^2-2T_CT_H-T_H^2).
 \label{eq:sm-witness}
\end{equation}
Thus $T_H/T_C<\sqrt2-1$ guarantees a nonzero response. The target is
entangled and the one-qubit gates have nonzero duration. This witness
establishes nonemptiness of the model class; it is not a claim that all
multilevel CZ implementations obey the effective diagonal model.

\section{Numerical illustrations and figure conventions}
\label{sm:numerical}
This section specifies the existing numerical studies reproduced in
the main-text figures and Fig.~\ref{fig:sm-mechanism}. They illustrate
the response mechanism; they are neither hardware measurements nor
an independent proof of the general error bounds. No new trajectories,
fits or sampled hardware outcomes are introduced here. Projected
sampling errors and numerical ensemble errors are distinguished below.

\subsection{Memory models and the theory illustration}
% Sources: engine/sq_model.py, memory(); tasks/T5_fig1_theory.py;
% results/T5_fig1_theory/summary.md and data.json.
The numerical $1/f$ memories use a smooth spectral regularization,
rather than the sharp frequency edges used for the finite-band
bounds in Sec.~\ref{sm:finite-band}. In the kick calculations the
unit-amplitude covariance is
\[
 C_g(\tau)=\frac{1}{\pi}\int_{g_{\rm lo}}^{g_{\rm hi}}
       \frac{e^{-g|\tau|}}{g}\,dg
 =\frac{E_1(g_{\rm lo}|\tau|)-E_1(g_{\rm hi}|\tau|)}{\pi},
 \qquad C_g(0)=\frac{\ln(g_{\rm hi}/g_{\rm lo})}{\pi},
\]
where $E_1$ is the exponential integral and $g$ is a decay rate.
This is a mixture of OU memories with approximately $1/f$
spectral behavior between its characteristic rates. The figure
labels ``IR'' or $1/\omega_{\rm IR}$ denote $g_{\rm lo}$ or its
inverse for this numerical model; they do not identify a hard
spectral edge. These curves illustrate the logarithmic-memory
mechanism, not an exact numerical realization of the sharp-edge
remainder bounds. Native-noise calculations discretize OU components
and represent components slower than their simulated rate grid by
an additional static variance.

Main-text Fig.~1(a) uses $g_{\rm lo}^{-1}=5\,\mu\mathrm{s}$
and $g_{\rm hi}^{-1}=25\,\mathrm{ns}$ to illustrate the memory
between blocks. Panel (b) instead uses
$g_{\rm lo}=10^{-11}\,\mathrm{ns}^{-1}$ and
$g_{\rm hi}^{-1}=25\,\mathrm{ns}$, a long-memory model chosen
for the theoretical comparison, not a measurement of a device
infrared cutoff. Its curves contract the complete second-order
kick kernel with this covariance. The untruncated propagation
uses $4000$ positive--negative pairs with matched sample covariance,
with the strength set at $\chi(13)=10^{-4}$; it is still a finite
numerical ensemble. For the contrast $y=2P-1$, both its bias and
its logarithmic coefficient are twice their survival-probability
counterparts. The normalized ratios in Fig.~1 use the same
observable convention in numerator and denominator.

The Richardson curves use the ordinary full-degree interpolation
weights on $\{1,3,\ldots,r_{\max}\}$; quadratic least squares
uses $\{1,r,r^2\}$ on those same nodes. The displayed PE\_log
sets end at $7$, $9$ or $11$. The horizontal coordinate
$\sum_k|w_k|$ is a sensitivity measure for those particular
intercepts, not physical noise amplification and not a general
finite-budget ordering. In this calculation, the full-response
Richardson bias at $(1,3,5)$ is $2.716$ in the plotted units;
the isolated $r^2\ln r$ moment in Sec.~\ref{sm:log-moment}
is $2.729\ldots$. They are different quantities: the former
includes the remaining second-order shape response.

\subsection{Single-qubit circuits and signals}
% Sources: T3 summary B--F; engine/exp1_circuits.json (sequences only).
% T3 overrides the legacy circuit-library duration with 25 ns.
All reference circuits contain nine rotations of angle $\pm\pi/2$,
each lasting $25\,\mathrm{ns}$, so $T=225\,\mathrm{ns}$.
Table~\ref{tab:sm-circuits} lists their chronological order.
The ideal target state is $U|0\rangle$, and $P$ is its survival
probability. Folding alternates the full sequence and its inverse.
An injected phase for a pulse is split equally between virtual-$Z$
rotations before and after that pulse. The phase sequence has the
declared temporal covariance and continues across folded blocks.
The single-qubit numerical model uses QuTiP-derived propagators;
relaxation and white dephasing act during pulses, while injected
phase kicks and continuous native noise are kept distinct.

\begin{table}[htbp]
\centering
\caption{Single-qubit circuit definitions used in the numerical
illustrations. Here $X_\pm=R_x(\pm\pi/2)$ and
$Y_\pm=R_y(\pm\pi/2)$. Orders are chronological, not a
right-to-left matrix product.}
\label{tab:sm-circuits}
\begin{tabular}{cl}
\toprule
Circuit & Pulse sequence \\
\midrule
C1 & $Y_-\,X_-\,X_-\,Y_-\,Y_-\,Y_-\,Y_-\,X_-\,Y_+$ \\
C2 & $Y_+\,X_+\,Y_+\,X_-\,Y_+\,X_-\,Y_+\,X_+\,Y_+$ \\
C3 & $Y_-\,X_-\,Y_-\,X_-\,Y_+\,X_-\,Y_+\,Y_+\,Y_+$ \\
C4 & $X_-\,Y_-\,Y_-\,X_-\,X_-\,X_-\,Y_+\,X_+\,Y_+$ \\
C5 & $Y_+\,Y_+\,X_-\,X_-\,Y_-\,Y_-\,X_+\,Y_+\,Y_+$ \\
C6 & $X_+\,X_+\,Y_+\,Y_+\,X_-\,Y_-\,X_-\,Y_+\,Y_+$ \\
C7 & $X_-\,Y_+\,X_-\,Y_-\,X_+\,Y_+\,X_-\,Y_-\,Y_-$ \\
C8 & $Y_+\,X_+\,Y_+\,X_-\,Y_-\,X_+\,Y_+\,X_-\,Y_-$ \\
\bottomrule
\end{tabular}
\end{table}

For the injection illustration the computed signal is
\[
 y(r)=\frac{2P_{\rm on}(r)-1}{2P_{\rm off}(r)-1},
 \qquad \chi(r)=-\ln y(r),\qquad
 m(r)=\frac{\chi(r)}{2\,\mathrm{amp}\,S(r)},
 \qquad S(r)=\frac{M_{0,\rm inj}r^2}{8}.
\]
The phase-rate covariance is $\mathrm{amp}\,C(t-t')$;
``on'' and ``off'' include and omit the additional injected
process, respectively. The C8 injection weight is
$M_{0,\rm inj}=22500\,\mathrm{ns}^2$. A noise process acting
continuously during the same control pulses instead has
$M_{0,\rm nat}=30737\,\mathrm{ns}^2$ in the separate native model.
These weights cannot be interchanged. The signal $m$ normalizes
by a nonzero frozen response, rather than identifying the full
convolution with a point value of $C(rT)$.
The zero-weight circuit C1 is used only for the unnormalized
weight comparison.

\begin{figure}[htbp]
\centering
\makebox[\textwidth][c]{\includegraphics[width=7in]{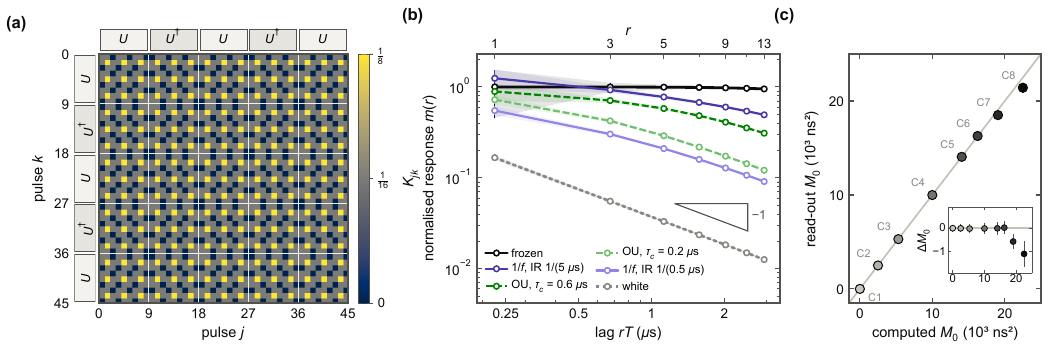}}
\caption{\textbf{The numerical response resolves temporal memory
and its circuit weighting.}
(a) C8 pulse-pair kernel at $r=5$ in the symmetric-kick model.
Here the displayed symmetric discrete kernel is
$K_{jk}=-(1/2)\partial_{\phi_j}\partial_{\phi_k}P|_{\phi=0}$,
so the second-order loss is
$1-P=\sum_{jk}K_{jk}\langle\phi_j\phi_k\rangle$.
It is a discrete survival-loss convention, not a direct sample
of the ordered operator kernel in the general theorem.
(b) Normalized finite-strength responses for the six memory models.
Curves and markers are predictions; bands and bars project one
sampling standard deviation at $4\times10^4$ samples per node
and on/off type. The white-noise guide has slope $-1$.
(c) Frozen-injection weight readout versus the calculated weight
for C1--C8, including a residual inset. C1 is a zero-response control;
the finite-strength C8 deviation includes higher orders.
Projected errors use $10^5$ samples per on/off type for C1--C7;
C8 retains the budget of (b). No hardware outcomes are plotted.}
\label{fig:sm-mechanism}
\end{figure}

Figure~\ref{fig:sm-mechanism} displays the pulse-pair weighting and
the six reference memories: uncorrelated kicks, frozen noise, OU
noise with $\tau_c=0.2$ and $0.6\,\mu\mathrm{s}$, and the smooth-band
$1/f$ models with $g_{\rm lo}^{-1}=0.5$ and $5\,\mu\mathrm{s}$.
The reference strength is $\chi(13)=0.4$. Under frozen injection
the plotted weight readout is
\[
 M_{0,\rm read}=\frac{8\chi(13)}{2\langle\delta^2\rangle\,13^2}.
\]
Its finite-strength deviation from the static coefficient is retained;
it is not measurement noise. The correction example in main-text
Fig.~2 instead uses $\chi(7)=0.45$ and the odd nodes through $9$.
Its central curves come from conditional phase-ensemble calculations;
the reference moment-matched design contains $250$ antithetic pairs.

% Unmodified artwork: folding_1f_sim/figures/fig2_mechanism_v2.pdf.

\subsection{Estimators and projected uncertainty}
% Sources: make_fig3_v2/v4.py; T3 F; T10 summary and data.json.
Exponential extrapolation is implemented here as unweighted least
squares of $\ln y$ to $\{1,r\}$ at $(1,3,5)$ followed by
exponentiation. Richardson is the linear-domain quadratic
interpolant on $(1,3,5)$. Single-qubit PE\_log is the unweighted
four-coordinate fit on $(1,3,5,7,9)$; the main multiqubit estimate
uses $(1,3,5,7)$. The ``Qiskit default'' legend identifies the
exponential extrapolator's functional form and node choice,
not the Runtime solver, its fallback decisions, or its full
mitigation pipeline. No gate twirling is included in these circuits.

The single-qubit RMSE curves retain the numerical signal bias
and parametric sampling variation; total counts include both
on and off acquisition. Each method uses equal samples per
node on its own node set. Consequently, the open design markers
for three-node and five-node methods are not at equal total cost.
The fixed-budget curves, rather than those markers alone, allow
a common-budget comparison. Sampling projections condition on
the stated noise, phase ensemble and readout model; they do not
include additional device drift or calibration uncertainty.

\begin{table}[htbp]
\centering
\caption{PE\_log single-qubit predictions at $10^7$ total on/off
samples, with $10^6$ samples per node and type. Entries are percentages.
The numerical ensemble standard error $\sigma_{\rm MC}$ is separate
from the sampling standard deviation $\sigma_{\rm shot}$ and RMSE.
The final column is their quadrature combination, not RMSE.}
\label{tab:sm-projected-errors}
\begin{tabular}{lrrrrr}
\toprule
Memory & Bias & RMSE & $\sigma_{\rm shot}$ & $\sigma_{\rm MC}$ &
 $\sigma_{\rm combined}$ \\
\midrule
$1/f$, $g_{\rm lo}^{-1}=5\,\mu\mathrm{s}$ & $-0.26$ & $0.38$ & $0.27$ & $0.05$ & $0.27$ \\
OU, $\tau_c=0.6\,\mu\mathrm{s}$ & $-0.04$ & $0.27$ & $0.27$ & $0.12$ & $0.30$ \\
\bottomrule
\end{tabular}
\end{table}

Table~\ref{tab:sm-projected-errors} separates the uncertainty
quantities: both predicted biases are consistent with zero
within the design's statistical resolution, but neither is
identically zero. At the fitted nodes, the reported maximum
PE\_log residuals are $0.023\%$ for the $1/f$ example and
$0.06\%$ for OU. Main-text Fig.~2(a) shows the relative node
residual $100[\widehat y(r)/y(r)-1]$; its point at $r=0$
instead shows $100[\widehat y(0)-1]$, since the ideal ratio is
one. The colored residual fill is not an uncertainty interval.

\subsection{Ten-qubit model and coefficient comparison}
% Sources: T2 summary and tasks/T2_tenqubit_qz01.py, one_setting;
% T11 summary, collapse; engine/devsim.py, kicked_ising.
The ten-qubit example uses a QZ01 calibration snapshot dated
September 28, 2026, for the contiguous chain q6--q15.
This is a numerical parameterization, not an experiment performed
for this manuscript. Each kicked-Ising step applies
$R_x(\pi/2)$ to every qubit and then even- and odd-bond
interactions $R_{ZZ}(-\pi/2)$. Each interaction is represented
by CZ and virtual $R_z(-\pi/2)$ on its two qubits, up to global
phase. The schematic in main-text Fig.~3(a) omits these virtual
phases for compactness. The model uses $25\,\mathrm{ns}$
single-qubit pulses and an assumed $40\,\mathrm{ns}$ CZ duration,
not a measured CZ timing. One to three steps have durations
$105$, $210$ and $315\,\mathrm{ns}$ and contain $9$, $18$ and
$27$ CZ gates, respectively. Complete global folding includes
all these operations and their inverse phases.

QuTiP is used for the Markovian density-matrix response and the
ideal control matrices. The classical colored-noise response
is averaged separately, with $10$ batches of $300$ antithetic
pairs in the target-observable study. The displayed total
expectation is assembled as
$E(r)=E_{\rm Markov}(r)E_{\rm classical}(r)$ for targets with
$E^*=1$. This is a factorized model for the joint response;
it is not a direct simultaneous mixed-noise master-equation
evolution for every trajectory, nor an exact factorization
theorem for general noise. The Markovian channels are placed
symmetrically around the gates in this model. Cross-mechanism
terms remain part of the general remainder derived earlier.

The main-text target example uses signed Pauli stabilizers.
The coefficient calculation is independent of the extrapolation
fit and uses $\theta=\sum_k A_k s_{2,k}/\pi$.
The compatible-observable study supplies $31$, $15$ and $7$
nontrivial stabilizers in the chosen measurement bases for one,
two and three steps. Reference and doubled total $1/f$ strength
give $106$ predictions per estimator. Its pooled exponential
slope is $116.6$ with $R^2=0.9972$; separate-setting slopes range
from $108.8$ to $125.6$ with $R^2\geq0.999$.
These are approximate model-dependent relationships. They do
not imply that every memory-induced bias equals a universal
multiple of $\theta$. No one of these signed-stabilizer targets
supplies a noise-insensitive null comparison in the analyzed
family; the illustration is a coefficient--bias comparison.

For that observable study, the largest PE\_log absolute bias
at each depth is $0.02\%$, $0.09\%$ and $0.50\%$ for the
reference model, and $0.05\%$, $0.47\%$ and $1.49\%$ for
doubled strength. Sampling bars use five acquired nodes,
$1,3,5,7,9$, with $2\times10^6$ samples per node; no counts
are reassigned to a method's smaller node subset. Observables
within one measurement basis share outcomes and hence have
correlated errors. In the depth--strength map, dots compare
absolute bias with projected shot and ensemble errors combined
in quadrature; boxes mark proposed follow-up settings. Neither
annotation denotes a measured pass/fail result.

\begin{table}[htbp]
\centering
\caption{Predicted signed target-observable bias (percent) in the
reference ten-qubit model. The entries are the rounded ensemble
means of the target-observable study, not finite-sample RMSEs.
They are distinct from the independently sampled compatible-
stabilizer ensemble in main-text Fig.~3(c).}
\label{tab:sm-target-estimators}
\begin{tabular}{lrrr}
\toprule
Estimator & One step & Two steps & Three steps \\
\midrule
Exponential, $(1,3,5)$ & $+0.78$ & $+1.38$ & $+3.28$ \\
Richardson, $(1,3,5)$ & $+0.06$ & $+0.17$ & $+0.45$ \\
PE\_log, $(1,3,5,7)$ & $+0.02$ & $-0.03$ & $-0.17$ \\
PE\_log, $(1,3,5,7,9)$ & $+0.04$ & $-0.06$ & $-0.35$ \\
\bottomrule
\end{tabular}
\end{table}

Richardson also has small predicted bias in the multiqubit
example (Table~\ref{tab:sm-target-estimators}). The example
therefore supports memory attribution and correction relative
to the specified exponential fit, not a universal performance
advantage over Richardson. Small residual bias also does not
establish the lower sampling variance of PE\_log.

\subsection{Numerical limits}
% Sources: T8 A/data.json; T6 A--B; confirmed author corrections.
The main-text single-qubit parameter map is an independently
sampled ensemble, not the ensemble used for its representative
response curve. On its sampled grid with $\chi(7)\leq0.45$,
$g_{\rm lo}^{-1}=2$--$10\,\mu\mathrm{s}$ and odd nodes through
$9$, the recorded PE\_log absolute bias is at most $0.31\%$.
This is a numerical result for that grid and ensemble, not a
continuous-domain bound. It is below the design's $2\sigma$
resolution of $0.54\%$; it does not establish that four
coordinates give an exact closure. The points at $\chi(7)=0.6$
are outside this quoted window, which does not itself mean
that the estimator fails there.

Short memories permit only partial correction: at
$g_{\rm lo}^{-1}=1\,\mu\mathrm{s}$ the map gives a
PE\_log bias of $+0.74\%$, and at
$\tau_c=0.4\,\mu\mathrm{s}$ for OU it gives $+0.99\%$.
At the shortest sampled memories the three methods share an
offset of about $+2\%$. Long memory is also not sufficient
for exact correction: at $\chi(7)=0.45$, the $1/f$ settings
$g_{\rm lo}^{-1}=20$ and $50\,\mu\mathrm{s}$ give
$-0.66\%$ and $-0.57\%$, respectively.

An independent strong-noise scan at
$g_{\rm lo}^{-1}=5\,\mu\mathrm{s}$ and $\chi(7)=1$
gives $-2.3\%$ with numerical ensemble error $0.5\%$;
the corresponding independently generated map cell is about
$-3.0\%$. These ensembles are not merged into a common bound.
A single slow telegraph fluctuator with correlation time
$5\,\mu\mathrm{s}$ gives $+4.1\%$, whereas the separate
five- and twenty-fluctuator examples give $+0.23\%$ and
$-0.17\%$. The single-fluctuator result falls outside the
Gaussian remainder assumptions. These examples make the
scope of the correction explicit without turning numerical
parameter choices into general hardware guarantees.

\end{document}